\documentclass[prd,amsmath,notitlepage,twocolumn]{revtex4-2}
\usepackage{graphicx}
\usepackage{float}
\usepackage{epstopdf,cancel}
\usepackage{epsf,latexsym,bbm,euscript}
\usepackage{amssymb,amsmath}
\usepackage{mathtools} 
\usepackage{times,graphics}
\usepackage{soul,xcolor}
\usepackage{mathtools}

\usepackage{url,hyperref}
\hypersetup{colorlinks,linkcolor={blue!55!black},citecolor={red!50!black},urlcolor={blue!45!black},breaklinks=true}

\usepackage{scalerel}
\usepackage{tikz}
\usetikzlibrary{svg.path}
\definecolor{orcidlogocol}{HTML}{A6CE39}

\def\6{{\langle}}
\def\9{{\rangle}}

\newcommand{\eqdef}{=\vcentcolon}

\newcommand{\tcr}{\textcolor{red}}

\usepackage{scalerel}
\usepackage{tikz}
\usetikzlibrary{svg.path}
\definecolor{orcidlogocol}{HTML}{A6CE39}
\tikzset{
	orcidlogo/.pic={
		\fill[orcidlogocol] svg{M256,128c0,70.7-57.3,128-128,128C57.3,256,0,198.7,0,128C0,57.3,57.3,0,128,0C198.7,0,256,57.3,256,128z};
		\fill[white] svg{M86.3,186.2H70.9V79.1h15.4v48.4V186.2z}
		svg{M108.9,79.1h41.6c39.6,0,57,28.3,57,53.6c0,27.5-21.5,53.6-56.8,53.6h-41.8V79.1z M124.3,172.4h24.5c34.9,0,42.9-26.5,42.9-39.7c0-21.5-13.7-39.7-43.7-39.7h-23.7V172.4z}
		svg{M88.7,56.8c0,5.5-4.5,10.1-10.1,10.1c-5.6,0-10.1-4.6-10.1-10.1c0-5.6,4.5-10.1,10.1-10.1C84.2,46.7,88.7,51.3,88.7,56.8z};
	}
}

\newcommand\orcidlink[1]{\href{https://orcid.org/#1}{\mbox{\scalerel*{
				\begin{tikzpicture}[yscale=-1,transform shape]
					\pic{orcidlogo};
			\end{tikzpicture}}{X}}}}

\newcommand{\be}{\begin{equation}}
\newcommand{\ee}{\end{equation}}
\newcommand{\ba}{\begin{eqnarray}}
\newcommand{\ea}{\end{eqnarray}}

\usepackage[caption=false]{subfig}
\newcommand{\subfigimg}[3][,]{%
  \setbox1=\hbox{\includegraphics[#1]{#3}}
  \leavevmode\rlap{\usebox1}
  \rlap{\hspace*{-10pt}\raisebox{.5\baselineskip}{\small{#2}}}
  \phantom{\usebox1}
}

\begin{document}

\title{The Hawking temperature of dynamical black holes via conformal transformations}

\author {Pravin Kumar Dahal\,\orcidlink{0000-0003-3082-7853}}
\email{pravin-kumar.dahal@hdr.mq.edu.au}

\author{Swayamsiddha Maharana\,\orcidlink{0009-0004-6006-8637}}
\email{swayamsiddha.maharana@hdr.mq.edu.au}

\affiliation{School of Mathematical and Physical Sciences, Macquarie University, NSW 2109, Australia}

\begin{abstract}

In this second part of our two-series on extracting the Hawking temperature of dynamical black holes, we focus into spacetimes that are conformal transformations of static spacetimes. Our previous investigation builds upon the Unruh-Hawking analogy, which relates the spacetime of a uniformly accelerating observer to the near-horizon region of a black hole, to obtain the Hawking temperature. However, in this work, we explicitly compute the Bogoliubov coefficients associated with incoming and outgoing modes, which not only yields the temperature but also thermal spectrum of particles emitted by a black hole. For illustration, we take the simplest nontrivial example of the linear Vaidya spacetime, which is conformal to the static metric and using this property, we analytically solve the massless scalar field in its background. This allows the explicit computations of the Bogoliubov coefficients to study the particle production in this spacetime. We also derive an expression for the total mass of such dynamical spacetimes using the conformal Killing vector. We then perform differential variations of the mass formula to determine whether the laws of dynamical black hole mechanics correspond to the laws of thermodynamics.

\end{abstract}

\maketitle

\section{Introduction}

Black holes have long captivated human imagination and their enigmatic nature has propelled them to the forefront of theoretical physics. They contain a strong-gravity regime where curvature of spacetime becomes sufficiently strong, offering a unique laboratory to explore the frontiers of physics~\cite{bambi:2017,EventHorizonTelescope:2020,LIGOScientific:2016}, particularly the elusive unification of gravity and quantum mechanics.

Contrary to the initial understanding that black holes could be completely characterized solely by a few macroscopic parameters, the discovery of their thermodynamic nature~\cite{Bardeen:1973,wald:2001,Carlip:2014} suggests they might possess an underlying microscopic structure~\cite{Hawking:1979,Strominger:1996,Ikeda:2021,Bombelli:1986,tHooft:1984}. This revelation caused a paradigm shift in perspective, opening research direction towards the hidden microscopic structure that underpins their thermodynamic properties, including the nature of Hawking radiation and other theoretical constructs.

The discovery of Hawking radiation not only completed black hole thermodynamics but also offered a crucial initial step towards reconciling gravity with quantum mechanics~\cite{Hawking:1975,Birrell:1982,Parker:2009}. Black hole can be viewed as an excited state of the gravitational field and thus should decay quantum mechanically. Quantum fluctuations of the metric allow energy to tunnel out of the black hole potential well, a process analogous to particle creation near deep potential wells in flat spacetime. The derivation of Hawking radiation is performed using the now well-established quantum field theory in curved spacetime~\cite{Birrell:1982,Parker:2009,Jacobson:2005}. In this framework, spacetime metric is treated classically but is coupled to matter fields which are treated quantum mechanically, and provides an excellent approximation beyond Planck scales. While quantum field theory in curved spacetime presents challenges in interpreting field operators in terms of creation and annihilation operators, this apparent drawback can be turned into an advantage. Vacuum state for an observer at early times, evolves into a thermal state populated with particles at late times~\cite{fabbri:2005,Birrell:1982,Parker:2009}.

Despite significant advancements in black hole physics since the discovery of both the black hole thermodynamics and Hawking radiation, the quest for a unified theory of quantum gravity remains a formidable challenge (although pre-Hawking radiation in the background of collapsing domain wall is discussed in Ref.~\cite{Vachaspati:2006ki}). We even lack an exact derivation of Hawking effects in dynamical spacetimes. We also lack the generalization of the laws of thermodynamics to dynamical spacetimes, rooted in first principles. In search of generalizing black hole mechanics to dynamical cases, various definitions of horizon~\cite{dahal:2022,Ashtekar:2004,Faraoni:2015} and surface gravity~\cite{Faraoni:2015,csv3,hw5,dahal:2023} have been proposed. There is still a debate on which horizon possesses thermodynamic characteristics and which surface gravity relates directly to Hawking temperature. Furthermore, each emitted Hawking particle contributes back to its source, inducing backreaction effects that continuously alter the spacetime dynamics. Some dynamical spacetimes are themselves solutions to the semiclassical backreaction problem, like the linear Vaidya metric at initial stages of macroscopic black hole evaporation~\cite{pf1,bardeen1981}. This inherent self-consistency distinguishes the derivation of particle production in such dynamic backgrounds from the static ones.

Therefore, these tasks are crucial for addressing conceptual issues and exploring new frontiers in black hole physics. This article is an attempt to tackle these tasks for a specific class of dynamical spacetimes, which are conformal to static ones. This represents the simplest, yet non-trivial, generalization, presenting an opportunity for uncovering novel aspects of particle production and thermodynamics that are not present in static backgrounds. This work is the continuation of our prior work~\cite{pf1}, where we employed a specific coordinate transformation, valid near the horizon, that allowed us to exploit the relationship between field theories defined on conformally related geometries to determine the temperature of Hawking radiation in dynamical spacetimes. Notably, this approach is akin to past derivation of the Hawking temperature of the static Schwarzschild spacetime, where the near-horizon expansion yields a simple Rindler geometry. For the Vaidya metric, the near-horizon expansion yields a conformal Rindler geometry, suggesting a formal relationship between the Hawking and Unruh effects despite their distinct origins. While our earlier study focused solely on temperature calculation, we now aim to formulate a comprehensive thermodynamic description of specific class of dynamic metrics.

This article is organized as follows: we address the problem of particle production in the linear Vaidya background in Sec.~\ref{s1}. For this, we first describe the relevant geometric features of linear Vaidya spacetime in Sec.~\ref{s1a}. We then solve the wave equation for the conformally coupled scalar field in its background in Sec.~\ref{s1b}. We use this incoming and outgoing wave solutions to extract Bogoliubov coefficients for calculating the spectrum of emitted particles in the linear Vaidya spacetime in Sec.~\ref{s1c}. In Sec.~\ref{s2}, we study the thermodynamic nature of dynamical spacetimes that are conformal to the static ones. For this, we take the conformal Killing vector to derive the formula for the total mass in such spacetimes, analogously to static spacetimes~\cite{Bardeen:1973}. Then, we perform the differential variation of this mass to see if we can obtain a relation that resembles the laws of thermodynamics in Sec.~\ref{s2a}. We discuss our results and conclude this article in Sec.~\ref{discus}.

Here, we consider a spacetime with metric $g_{\mu\nu}$ of signature $(-,+,+,+)$. A timelike unit vector ${\hat T}^\mu$ is normal to the volume $\Sigma$, while a spacelike unit vector $N^\mu$ is  normal to the surface $\partial\Sigma$. Both vectors become null at the event horizon $\partial\Sigma_{+}$. Furthermore, $\left(U^\mu, V^\mu\right)$ represent orthonormal vectors tangent to the surface $\partial\Sigma$. Some notations employed here aligns with Ref.~\cite{wb13}, and the curvature convention follows that of Ref.~\cite{MTW:73}. We set $\hbar=k_B=G=c=1$ throughout the analysis.

\section{Particle production in Linear Vaidya} \label{s1}

In this section, we present the exact calculation of Hawking radiation in dynamical spacetimes, which are conformal transformations of static spacetimes. For this purpose, we consider the specific example of linear Vaidya spacetime in advanced coordinates, which serves as the simplest nontrivial black hole solution of the semiclassical self-consistent approach~\cite{mann:2022,Dahal:2023suw} and also a model of an evaporating black hole, that approximately captures backreaction effects~\cite{bardeen1981,pf1}.

\subsection{Relevant geometric properties of linear Vaidya} \label{s1a}

Evaporating Vaidya black hole, in advanced coordinates, takes the form
\begin{equation}
    ds^2= - f(v,r) dv^2+ 2 dv dr+ r^2 d\Omega^2, \quad f(v,r)= 1- \frac{2 M(v)}{r}. \label{v1}
\end{equation}
Refs.~\cite{mann:2022,Dahal:2023suw} have shown that this metric constitute a self consistent solution of the semiclassical Einstein equations. Furthermore, we have also shown that this spacetime is an excellent near horizon approximation of the generic self-consistent spherically symmetric black hole solutions~\cite{Dahal:2023suw,pf1}. For linearly decreasing mass $2 M(v)= r_0- \alpha v$, this metric would be conformal to the static metric in advanced coordinates (see appendix~\ref{app1})
\begin{multline}
    ds^2= e^{-2\alpha {\cal T}/r_0} \Bigg(-\left(1-\frac{r_0}{R}+ \frac{2\alpha R}{r_0}\right) dT^2+ \\
    \frac{1}{1-\frac{r_0}{R}+ \frac{2\alpha R}{r_0}} dR^2+ R^2 d\Omega^2\Bigg). \label{cs3}
\end{multline}
Solutions of the equation
\begin{equation}
    1-\frac{r_0}{R}+ \frac{2\alpha R}{r_0}=0,
\end{equation}
gives locations of the horizon of the corresponding static spacetime
\begin{equation}
    r_{+}= \frac{-1+ \sqrt{1+ 8 \alpha}}{2 \alpha} M(v), \quad r_{-}= -\frac{1+\sqrt{1+8\alpha}}{2\alpha} M(v). \label{hor5}
\end{equation}
Here, $r_+$ denotes outer and $r_-$ denotes inner horizons. Since the event horizon remains invariant under conformal transformations, $r_+$ is also the event horizon of linear Vaidya spacetime. In $(T,R)$ coordinates, its location is time independent despite the dynamical nature of the metric. This is what we expect from the global nature of the event horizon, which necessitates understanding the entire spacetime history to determine its location.

\begin{figure}[!htbp]
\includegraphics[width=0.48\textwidth]{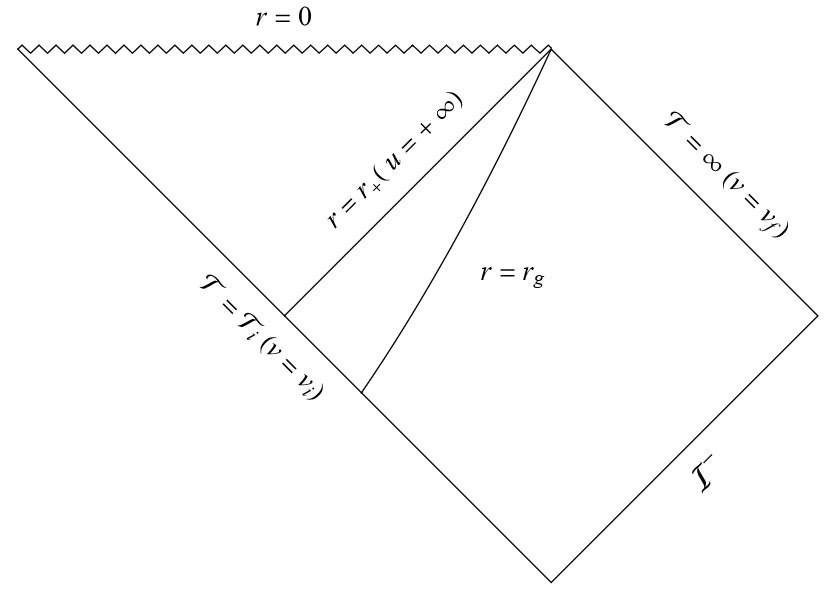}
\caption{Carter–Penrose diagram of the evaporating Vaidya spacetime. ${\mathcal I}^-$ represents the past null infinity. $v_i$ and $v_f$ denote arbitrary initial and final times, respectively. $u=\infty$ correspond to the event horizon and $r=r_g$ gives an apparent horizon (timelike membranes). The region between the event and apparent horizons is the quantum ergosphere, where an analog of Penrose process could occur in spherical symmetry. $r=0$ is the singularity of Vaidya spacetime. Various studies suggest the plausibility of naked singularity in Vaidya spacetime~\cite{Mkenyeleye:2014,Vertogradov:2016}. However, we do not consider formation history here, because of which, concerns about the appearance of such singularities are not relevant for us.}
	\label{f1}
\end{figure}

Dynamical spacetimes conformal to static spacetimes can be described by the line element of Eq.~\eqref{come1}.
One can also introduce the notion of a conformal Killing vector in such spacetimes, which is the Killing vector in the corresponding conformal static spacetime. It satisfies the conformal Killing equation
\begin{equation}
    K_{\nu;\mu}+ K_{\mu;\nu}= 2 {\cal B} g_{\mu\nu}, \label{ckv6}
\end{equation}
for some arbitrary function ${\cal B}$. For linear Vaidya metric
\begin{equation}
    K^\mu= -\left(\frac{M}{M'}, r,0,0\right)= \left(\frac{r_0}{\alpha}, 0,0,0\right), \label{ckv7}
\end{equation}
where the second equality is in the advanced $(v,r)$ and the third equality is in static $(T,R)$ coordinates and $\mathcal{B} = -1$. The existence of the conformal Killing vector allows the definition of another notion of horizon, known as conformal Killing horizon. It is the two-surface where the conformal Killing vector becomes null, and coincides with the event horizon when both horizons exist. Throughout this article, we will use both terms interchangeably. Finally, we introduce the notion of an apparent horizon, which represents the outermost boundary of the trapped region (a region of spacetime where both the outgoing and ingoing null geodesic expansions are negative). Unlike the event and conformal Killing horizons, the apparent horizon is not invariant under the conformal transformations. For the Vaidya metric of Eq.~\eqref{v1}, it is located at $r= 2 M(v)$. The Carter–Penrose diagram of the evaporating Vaidya spacetime, showing the locations of the respective horizons, is depicted in Fig.~\ref{f1}.

\subsection{Scalar field on linear Vaidya} \label{s1b}

In general spherically symmetric spacetimes, there exists a preferred class of fiducial observers that allows one to define a geometric quantity known as the Kodama vector~\cite{av15}. This vector is timelike outside the apparent horizon and spacelike inside it. So, there exists a region of spacetime outside the event and inside the apparent horizon, known as the quantum ergosphere, where some physical four-momenta can have negative energy. Here, the energy is defined analogously as $E= \chi\cdot p$, where $\chi$ represents the Kodama vector and $p$ represents the four-momentum of a particle, although this quantity is not conserved as in the stationary case. Consequently, the associated mechanism of the classical Penrose process~\cite{Penrose:1971,sc15} manifests in the context of spherically symmetric dynamical black holes. Consider a particle with energy $E_0>0$ entering the quantum ergoregion. Upon entering, the particle breaks up into two with energies $E_1$ and $E_2$, such that, $E_0= E_1+ E_2$. If the energy absorbed by the black hole is negative $E_1<0$, then the energy of the particle coming out from it is greater than the energy of the initial ingoing particle ( $E_2>E_0$). This excess energy comes from the mass loss of the black hole. Further investigations into the Penrose process within the quantum ergosphere are reserved for future studies.

We now investigate the problem of particle creation in the linear Vaidya spacetime. For this, we take a free massless scalar field $\phi$ conformally coupled to spacetime curvature
\begin{equation}
    \phi_{;\mu\nu} g^{\mu\nu}+ \frac{1}{6} R^\mu_\mu \phi= 0, \label{fe6}
\end{equation}
where $R^\mu_\mu$ denotes the Ricci scalar, semicolon ($;$) denotes the covariant derivative and the factor $1/6$ corresponds to the conformal coupling. Conformally coupled field equation is invariant under the conformal transformations of the form~\cite{Parker:2009}
\begin{align}
    {\tilde g}_{\mu\nu}= \Omega^2(x) g_{\mu\nu}, \quad {\tilde \phi}= \Omega^{-1}(x) \phi.
\end{align}
So, to find the explicit solution of the field equation in the linear Vaidya background, we find the solution in the corresponding conformal static spacetime. Then the solution in the linear Vaidya spacetime can be obtained from the simple rescaling
\begin{equation}
    \phi= e^{\alpha {\cal T}/r_0} {\tilde \phi},
\end{equation}
where we have substituted the explicit value of the scale factor $\Omega$ from Eq.~\eqref{cs3}. To solve the wave equation~\eqref{fe6} in the conformal static spacetime, let us first expand it as
\begin{equation}
    \partial_\mu \left( |g|^{1/2} g^{\mu\nu} \partial_\nu{\tilde\phi} \right)+ \frac{|g|^{1/2}}{6} R^\mu_\mu {\tilde\phi}= 0.
\end{equation}
This equation can be solved by the separation of variables and the solution corresponding to the angular variables is spherical harmonics $Y_{l,m}(\theta,\varphi)$ (as we are considering the spherically symmetric background). Moreover, the solution corresponding to the temporal variable is $e^{\pm i k T}$, for some nonnegative constant $k$. Consequently, only an ordinary differential equation in radial variable $R$ remains, which has the form
\begin{multline}
    \partial_R \left( \left(1-\frac{r_0}{R}+ \frac{2\alpha R}{r_0} \right) R^2 \partial_R {\tilde\phi}_R \right)+ \\
    \left( \frac{k^2 R^2}{1-r_0/R+ 2\alpha R/r_0}- l(l+1) \right){\tilde\phi}_R=0,
\end{multline}
where ${\tilde\phi}_R$ is the component of ${\tilde\phi}$, depending only on coordinate $R$. Now, for convenience, we define the tortoise coordinate $R^*$ for the static metric as
\begin{equation}
    dR^*= \frac{1}{1-r_0/R+ 2\alpha R/r_0} dR,
\end{equation}
such that, the above equation becomes
\begin{multline}
    \frac{d^2\tilde\phi_R}{d{R^*}^2}+ \frac{2}{R} \left(1-\frac{r_0}{R}+ \frac{2\alpha R}{r_0} \right) \frac{d{\tilde\phi}_R}{d{R^*}}+\\
    \left( k^2 - l(l+1) \frac{1-r_0/R+ 2\alpha R/r_0}{R^2} \right){\tilde\phi}_R=0. \label{de14}
\end{multline}
In appendix~\ref{appde2}, we have reduced this equation into the Heun differential equation. The solutions to this equation are denoted as ${\cal G}_{11}^\pm$ and ${\cal G}_{12}^\pm$.  The Heun differential equation has wide-ranging applications across various branches of physics~\cite{hortacsu:2011}, including the study of black hole perturbations and their quasi-normal modes in general relativity~\cite{borissov2009, giscard:2020}.

Field $\phi$ can be quanatized, for example, by canonical quantization. The generic solution of the quantum scalar field in the conformal static spacetime can be written as the linear superposition of individual solutions (constituting both incoming and outgoing waves of positive and negative frequencies)
\begin{equation}
    \tilde \phi= \int dk \left(A_k f_k^{\mathrm{in}}+ A_k^\dagger f_k^{*\mathrm{in}}+ B_k f_k^{\mathrm{out}}+ B_k^\dagger f_k^{*\mathrm{out}} \right),
\end{equation}
where
\begin{align}
    f_k^{\mathrm{in}}=& e^{-i k (T+ R^*)} \left(C_1 {\cal G}_{11}^- + C_2 {\cal G}_{12}^- \right)Y_{l,m}(\theta,\varphi), \label{inm16}\\
    f_k^{\mathrm{out}}=& e^{-i k (T- R^*)} \left(D_1 {\cal G}_{11}^+ + D_2 {\cal G}_{12}^+ \right) Y_{l,m}(\theta,\varphi), \label{ogm17}
\end{align}
and $f_k^{*\mathrm{in}}$ and $f_k^{*\mathrm{out}}$ are their complex conjugates (index $l,m$ are suppressed for convenience). Here, $A_k, B_k$ are time independent quantum operators and $A_k^\dagger, B_k^\dagger$ are their Hermitian conjugates.

\subsection{Particle production and its spectrum} \label{s1c}

After obtaining the general solution to the scalar field equation, we now turn to the problem of particle production in the linear Vaidya background. We proceed by explicit computation of the relevant Bogoliubov coefficients, which relate the ingoing and outgoing modes of positive and negative frequencies. As their computations require tracing null rays (which are conformally invariant) backward from the asymptotic region, we expect obtaining the Hawking temperature that is invariant under conformal transformations~\cite{jacobson1993d}.

As the linear Vaidya spacetime is asymptotically flat, we can construct a natural field quantization in the ``$\mathrm{in}$" and ``$\mathrm{out}$" regions. For the massless field, we are considering here, these regions can be regarded as past null infinity ${\mathcal I}^-$ and future null infinity ${\mathcal I}^+$, respectively. Let us first consider incoming modes only
\begin{equation}
    \begin{aligned}
        \phi^{\mathrm{in}}=& e^{\alpha {\cal T}/r_0} \tilde\phi^{\mathrm{in}}= e^{\alpha {\cal T}/r_0} \int_k dk \left(A_k f_k^{\mathrm{in}}+ A_k^\dagger f_k^{*\mathrm{in}}\right),\\
        f_k^{\mathrm{in}}=& e^{-i k {\cal T}} \left(C_1 {\cal G}_{11}^- + C_2 {\cal G}_{12}^- \right)Y_{l,m}(\theta,\varphi), \label{if31}
    \end{aligned}
\end{equation}
where we have used Eq.~\eqref{ac6} to write incoming modes in time coordinate ${\cal T}$. Moreover, we can use Eq.~\eqref{r3} to write incoming modes in advanced coordinate as
\begin{equation*}
    f_k^{\mathrm{in}}= \left(1- \frac{\alpha v}{r_0}\right)^{i k r_0/\alpha} \left(C_1 {\cal G}_{11}^- + C_2 {\cal G}_{12}^- \right)Y_{l,m}(\theta,\varphi).
\end{equation*}
Or,
\begin{equation}
    f_k^{\mathrm{in}}\bigg|_{r\to\infty}\approx \frac{1}{R} \left( C_1 R^{\frac{i k r_0}{\alpha r_+}} + C_2 \right) \left(1- \frac{\alpha v}{r_0}\right)^{\frac{i k r_0}{\alpha}} Y_{l,m}(\theta,\varphi).
\end{equation}
As the term with the coefficient $C_1$ does not give the desired asymptotic form, we exclude it from our solution to obtain incoming modes near past null infinity ${\mathcal I}^-$
\begin{equation}
    f_k^{\mathrm{in}}\bigg|_{r\to\infty}\approx \frac{C_2}{R} \left(1- \frac{\alpha v}{r_0}\right)^{i k r_0/\alpha} Y_{l,m}(\theta,\varphi) . \label{im20}
\end{equation}
$f_k^{\mathrm{in}}$ and $f_k^{*\mathrm{in}}$ constitute a complete set of solutions (along with $f_k^{\mathrm{out}}$ and $f_k^{*\mathrm{out}}$), satisfying orthonormality conditions on ${\mathcal I}^-$ (see appendix~\ref{appm4}). Moreover, $f_k^{\mathrm{in}}$ contains only positive frequencies with respect to the canonical affine parameter along the null geodesic generators of ${\mathcal I}^-$. Consequently, operators ${A_k}$ and ${A_k^\dagger}$ satisfy canonical commutation relations and have the natural interpretation as the annihilation and creation operators for incoming particles. Since massless fields are entirely determined by their data on ${\mathcal I}^-$, the field $\phi$ can be expressed in the form of Eq.~\eqref{if31} everywhere.

Now, we consider outgoing waves only
\begin{equation}
    \phi^{\mathrm{out}}= e^{\alpha {\cal T}/r_0} \tilde \phi^{\mathrm{out}}= e^{\alpha {\cal T}/r_0} \int_k dk \left(B_k f_k^{\mathrm{out}}+ B_k^\dagger f_k^{*\mathrm{out}}\right) ,
\end{equation}
where $f_k^{\mathrm{out}}$ is defined in Eq.~\eqref{ogm17}. Analogously, using the definition of the retarded coordinate (given in appendix~\ref{app1}), we can write outgoing modes as
\begin{align}
    f_k^{\mathrm{out}}=& e^{-i k u} \left(D_1 {\cal G}_{11}^+ + D_2 {\cal G}_{12}^+ \right) Y_{l,m}(\theta,\varphi) \nonumber\\
    \implies f_k^{\mathrm{out}}\bigg|_{r\to\infty}\approx & \frac{1}{R} \left( D_1 R^{-\frac{i k r_0}{\alpha r_+}} + D_2 \right) e^{-i k u} Y_{l,m}(\theta,\varphi) .
\end{align}
As the term with the coefficient $D_1$ does not give the desired asymptotic form, we exclude it from our solution to obtain outgoing modes near future null infinity ${\mathcal I}^+$
\begin{equation}
    f_k^{\mathrm{out}}\bigg|_{r\to\infty}\approx \frac{D_2}{R} e^{-i k u} Y_{l,m}(\theta,\varphi) . \label{im22}
\end{equation}
These purely outgoing waves have zero Cauchy data on the event horizon. It can be shown that the solution $f_k^{\mathrm{out}}$ and $f_k^{*\mathrm{out}}$ satisfy orthonormality conditions on ${\mathcal I}^+$ (appendix~\ref{appm4}). Moreover, $f_k^{\mathrm{out}}$ contain only positive frequencies with respect to the canonical affine parameter along the null geodesic generators of ${\mathcal I}^+$. Consequently, the operators $B_k$ and $B_k^\dagger$ can be interpreted as the annihilation and creation operators for outgoing particles, that is, for particles on ${\mathcal I}^+$.

As massless fields are completely determined by their data on ${\mathcal I}^-$, one can express $f_k^{\mathrm{out}}$ as linear combinations of $f_k^{\mathrm{in}}$ and $f_k^{*\mathrm{in}}$
\begin{equation}
    f_k^{\mathrm{out}}= \int dk' \left(\alpha_{kk'} f_{k'}^{\mathrm{in}}+ \beta_{kk'} f_{k'}^{*\mathrm{in}}\right),
\end{equation}
where
\begin{equation}
    \alpha_{kk'}= (f_{k'}^{\mathrm{in}},f_k^{\mathrm{out}}),\quad \beta_{kk'}= - (f_{k'}^{*\mathrm{in}},f_k^{\mathrm{out}}), \label{bc25}
\end{equation}
are complex numbers independent of coordinates.

\begin{figure*}[!tbp]
	\centering
	\vspace{-15mm}
	\begin{tabular}{@{\hspace*{-.05\linewidth}}p{0.15\linewidth}@{\hspace*{0.35\linewidth}}p{0.40\linewidth}@{}}
		\centering
		\subfigimg[scale=0.45]{(a)}{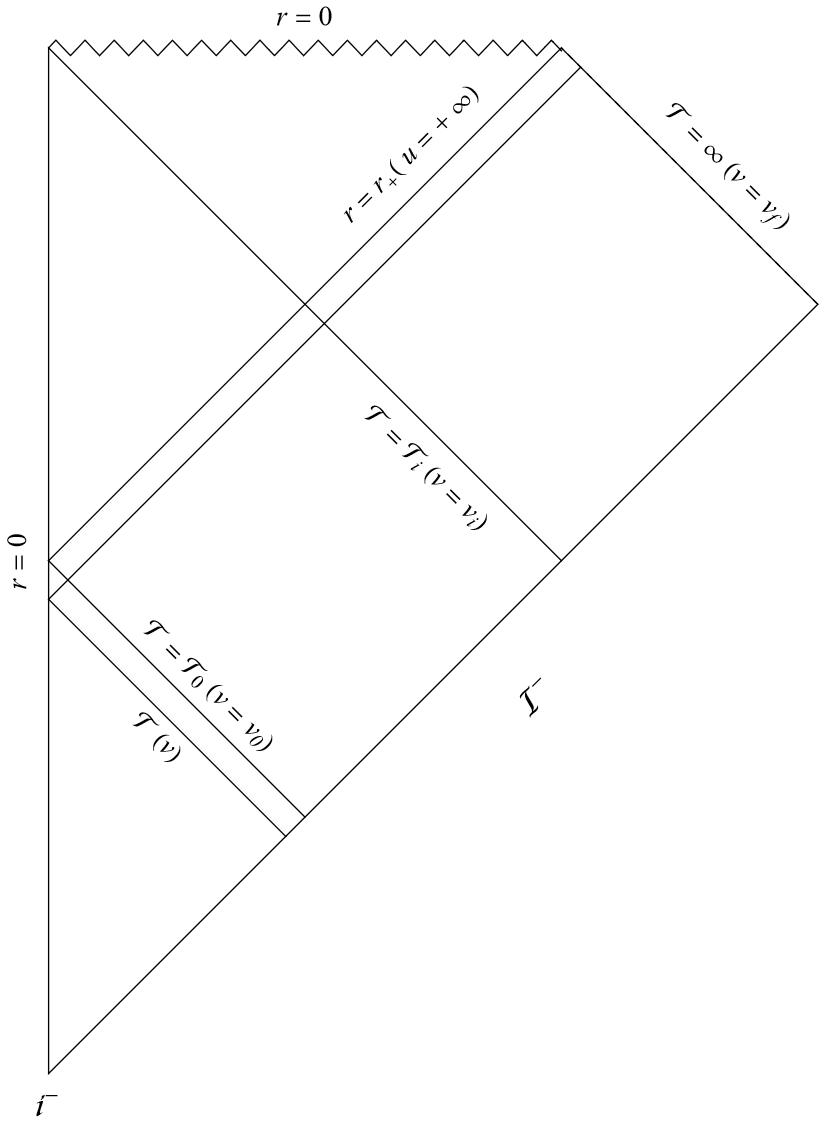} &
		\subfigimg[scale=0.50]{(b)}{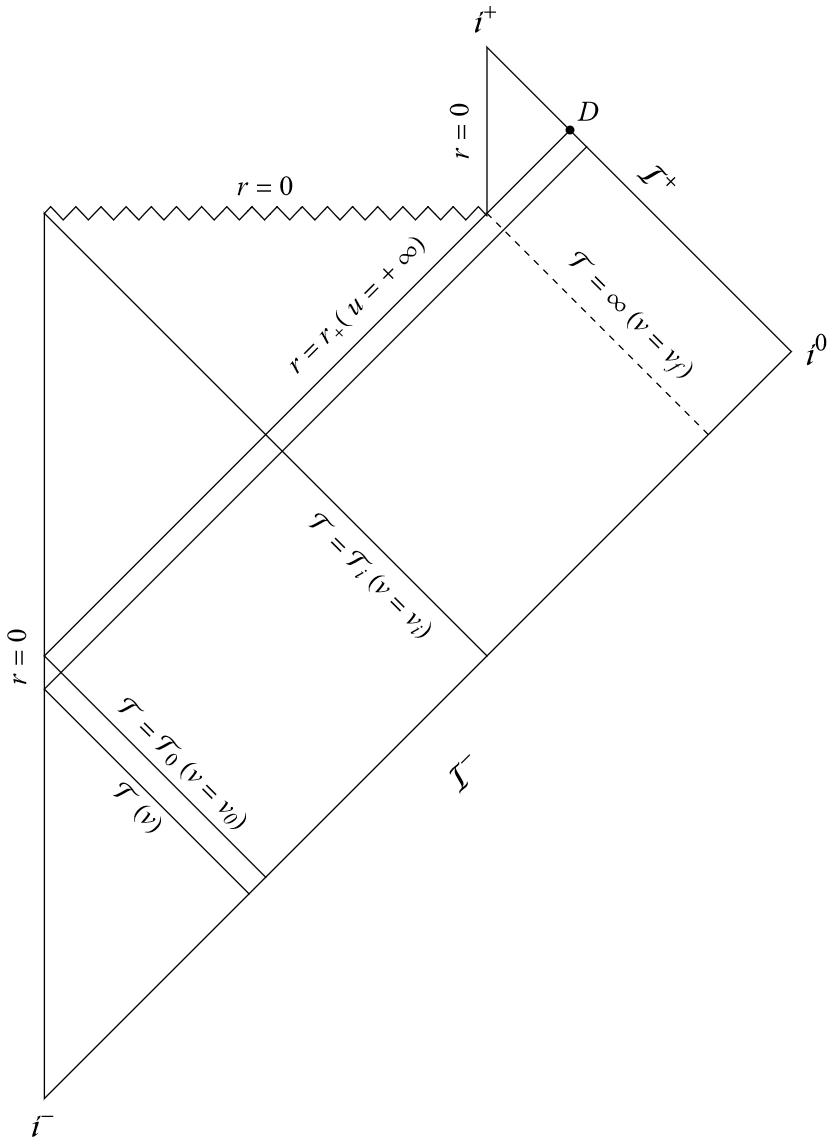}
	\end{tabular}
	\vspace{0mm}
	\caption{Carter–Penrose diagram of evaporating Vaidya spacetime a) Considering the formation phase only, and b) Considering both the formation and evaporation stages. Past and future timelike infinities are labelled by $i^-$ and $i^+$, respectively and past and future null infinities are denoted by ${\mathcal I}^-$ and ${\mathcal I}^+$, respectively. Spacelike infinity is represented by $i^0$ and $u=\infty$ corresponds to the event horizon. ${\cal T}(v)$ is the ray emerging in the distant future, passing through the centre of the collapsing body just before the formation of the event horizon and extending towards ${\mathcal I}^-$. This ray, in figure (a) terminates at ${\mathcal T}=\infty$, which corresponds to the future null infinity of the conformal static spacetime. As this situation is not different from the corresponding situation in the Schwarzschild case, the observer at ${\mathcal T}=\infty$ perceives a constant temperature. We need to rescale this temperature by some appropriate redshift factor to obtain the temparature of the Vaidya spacetime as perceived by an observer at ${\mathcal I}^+$. Here, $D$ denotes the point at which the black hole undergoes complete evaporation for this observer.}
	\label{fig2}
\end{figure*}

Understanding the phenomenon of particle creation resulting from the mixing of positive and negative frequencies necessitates the explicit calculation of Bogoliubov coefficients $\alpha_{kk'}$ and $\beta_{kk'}$. This requires considering the formation phase as well. However, we will focus on the asymptotic form of these coefficients, which remains independent of the details of collapse. For this purpose, we consider an outgoing wave $f_k^{\mathrm{out}}$ propagating backward from ${\mathcal I}^+$, just before the formation of an event horizon. A part $f_{k1}^{\mathrm{out}}$ of it will be scattered by the field outside the collapsing body and will end up on ${\mathcal I}^-$ with the same frequency. The remainder $f_{k2}^{\mathrm{out}}$ will enter the collapsing body, where it will be partly scattered and partly reflected through the centre, eventually emerging onto ${\mathcal I}^-$. In the asymptotic limit we are taking, this $f_{k2}^{\mathrm{out}}$ constitutes the dominant source of Hawking radiation. This is attributed to the accumulation of the surfaces of constant phase of solution $f_{k}^{\mathrm{out}}$ near the event horizon as the retarded time coordinate $u$ approaches infinity. To an observer on the collapsing body, the wave would appear to have an extremely high blueshift. On ${\mathcal I}^-$, $f_{k2}^{\mathrm{out}}$ would have infinite number of cycles just before the advanced time ${\cal T}= {\cal T}_0$, where ${\cal T}_0$ is the time of formation of an event horizon. Penrose diagram, depicting this ray is shown in Fig.~\ref{fig2}. As we are only interested in $f_{k2}^{\mathrm{out}}$, we will simply denote it as $f_{k}^{\mathrm{out}}$ (its phase is calculated in appendix~\ref{app2}).

Now, we have all the ingredients to calculate Bogoliubov coefficients of Eq.~\eqref{bc25}
\begin{align}
    \alpha_{kk'}=& C \int_{-\infty}^{{\cal T}_0} d{\cal T} \left(\frac{k'}{k}\right)^{1/2} e^{-i k u} e^{i k' {\cal T}} , \label{bc26}\\
    \beta_{kk'}=& C \int_{-\infty}^{{\cal T}_0} d{\cal T} \left(\frac{k'}{k}\right)^{1/2} e^{-i k u} e^{-i k' {\cal T}} \label{bc27},
\end{align}
where $C$ is constant. As we show in appendix~\ref{app2}, following relation holds between Bogoliubov coefficients $\alpha_{kk'}$ and $\beta_{kk'}$
\begin{equation}
    |\alpha_{kk'}|^2= e^{-\frac{\pi k (1- \sqrt{1+8\alpha}) r_0}{\alpha \sqrt{1+8\alpha} }} |\beta_{kk'}|^2. \label{bc28}
\end{equation}
Let us define the initial vacuum state $|0\rangle$, which is the state with no particles on past null infinity, as
\begin{equation}
    A_k |0\rangle= 0 \quad \textrm{for all} \quad k.
\end{equation}
However, the coefficients $\beta_{kk'}$ do not vanishes in general, due to which the initial vacuum state will not appear as a vacuum state to an observer at future null infinity. Expectation value of the number operator for $k^{th}$ outgoing mode measured by an observer at some late time is
\begin{equation}
    \langle 0_-|B_k^\dagger B_k|0_-\rangle= \sum_{k'} |\beta_{kk'}|^2.
\end{equation}
To evaluate this, we consider the outgoing wave $f_k^{\mathrm{out}}$ propagating backward from future null infinity, labeled as ${\cal T}(v)$ in Fig.~\ref{fig2}. A fraction $1-\Gamma_k$ of this wave will be scattered by the Vaidya background and a fraction $\Gamma_k$ will enter the collapsing body
\begin{equation}
    \Gamma_k= \int_0^\infty \left( |\alpha_{kk'}|^2- |\beta_{kk'}|^2\right) dk.
\end{equation}
The negative sign in front of the second integrand accounts for negative frequency components of $f_{k2}^{\mathrm{out}}$ making a negative contribution to the flux into the collapsing body. So, at late times, the number of created particles per unit time and per unit angular frequency that passes through some surface which is much larger than the circumference of the black hole event horizon is
\begin{equation}
    (2\pi)^{-1} \Gamma_k \left(\exp\left(-\frac{\pi k (1- \sqrt{1+8\alpha}) r_0}{\alpha \sqrt{1+8\alpha} }\right)-1\right)^{-1}.
\end{equation}
Thus, the Vaidya black hole exhibits emission and absorption behavior analogous to a gray body of absorptivity $\Gamma(k)$ and temperature $T$
\begin{equation}
    T= -\frac{\alpha \sqrt{1+ 8\alpha}}{\pi (1- \sqrt{1+8\alpha}) r_0}= \frac{r_+- r_-}{4\pi r_+ (r_++ r_-)}= \frac{\kappa}{2 \pi}. \label{gsg33}
\end{equation}
Here, the second equality relates the temperature to the inner and outer horizons defined in Eq.~\eqref{hor5} (this form is valid for the Kerr spacetime also). This constant surface gravity we have recovered is the geometric surface gravity~\cite{jacobson1993d}, which is invariant under conformal transformations. The invariance under conformal transformations arises from the geometric optics approximation, which requires tracing null rays (conformally invariant) only with infinitely high blueshift backward from the asymptotic region. It is clear from Fig.~\ref{fig2} that the observer at ${\cal T}=\infty$ measures a constant temperature as the dominant contribution to Hawking radiation come from rays with infinitely large retarded time $u$ (where geometric optics approximation holds). However, this is not necessarily true for the observer at ${\mathcal I}^+$. Below, we perform some normalizations and rescaling of a frame to obtain the temperature perceived by an observer at infinity.

\section{thermodynamics} \label{s2}

We now turn our attention to the thermodynamics of such spacetimes which are conformal to static ones. Before proceeding, however, some important points merit discussion. For the specific case of linear Vaidya metric, we consider the regime where the dynamics of the spacetime solely arise from the backreaction effects of Hawking evaporation. Our prior work has demonstrated the existence of such a regime for the Vaidya metric~\cite{pf1,Dahal:2023suw}. Within this regime, where we treat a black hole as a purely thermodynamic system, the first law of black hole thermodynamics is expected to hold~\cite{Hayward:1997jp}
\begin{equation}
    dM= T dS/4,
\end{equation}
where $T$ denotes the black hole temperature. Additionally, it is natural to expect $S$ to be the horizon area and $M$ to be the horizon's mass. Assuming $r_g$ as the location of the apparent horizon, its area is $4\pi r_g^2$ and the Misner-Sharp mass of the black hole is $r_g/2$. We can therefore calculate the temperature of a black hole from the this equation as
\begin{equation}
    T= 4 \frac{dM}{dS}= \frac{1}{4\pi r_g}.
\end{equation}
As previously mentioned, this temperature is perceived by an observer at infinity and can be obtained by rescaling the constant temperature we obtained in Eq.~\eqref{gsg33} by the prescription given in Refs.~\cite{nielsen2013a,Dicke:1961}
\begin{equation}
    \Omega^{-1} \xi \frac{\kappa}{2\pi} \approx \frac{1}{8\pi M}, \label{t35}
\end{equation}
where, $\Omega$ is the conformal factor and $\xi$ accounts for the appropriate normalization of the conformal Killing vector.

However, our analysis of particle production is based on the nature of quantum fields near an event horizon, not an apparent horizon. Therefore, we expect the event horizon (and not the apparent horizon) to appear explicitly in the expression of the first law. However, in static $(T,R)$ coordinates, 1) the notion of the event horizon is constant and 2) it depends on the choice of initial time (it is explicit in Eq.~\eqref{hor5} that it depends on $r_0$).
We can rescale the time variable to chose an arbitrary initial mass $r_0/2$ such that the Misner-Sharp mass remains unchanged,
\begin{equation}
    2 M= r_0- \alpha v= r_1- \alpha \left(v- \frac{r_0-r_1}{\alpha}\right).
\end{equation}
Thus, a static system could be obtained by a time translation. This translation is facilitated by the fact that the linearly evaporating Vaidya black hole is a conformal transformation of some static spacetime.


The mass formulae for dynamical  spacetimes conformal to static spacetimes can be derived by exploiting the property of the conformal Killing vector given in Eq.~\eqref{ckv6}. Contracting this equation by the metric tensor yields
\begin{equation}
    2 K^\mu_{~~;\mu}= 2 {\cal B} g^\mu_{~~\mu}. \label{sg40}
\end{equation}
Applying the relation for the commutation of covariant derivatives
\begin{equation}
    A^\mu_{~~;\mu\beta}- A^\mu_{~~;\beta\mu}= - R_{\nu\beta} A^\nu,
\end{equation}
to the conformal Killing vector $K^\mu$ gives (here, $A^\mu$ is an arbitrary vector)
\begin{equation}
    K^\mu_{~~;\beta\mu}= R_{\nu\beta} K^\nu+ \left({\cal B} g^\mu_{~~\mu}\right)_{,\beta}.
\end{equation}
Integrating this equation over some spacelike hypersurface $\Sigma$ gives
\begin{multline}
    \int_\Sigma K^\mu_{~~;\beta\mu} {\hat T}^\beta \sqrt{|\gamma^{(\Sigma)}|} d^3 x= \\
    \int_\Sigma \left(R_{\nu\beta} K^\nu +\left({\cal B} g^\mu_{~~\mu}\right)_{,\beta} \right){\hat T}^\beta \sqrt{|\gamma^{(\Sigma)}|} d^3 x,
\end{multline}
where $\gamma^{(\Sigma)}$ is the induced metric and ${\hat T}^\beta$ is a unit timelike vector normal to $\Sigma$. As the quantity $\left(K_{\mu;\beta}- {\cal B} g_{\mu\beta} \right)$ is antisymmetric (see Eq.~\eqref{ckv6}), we can use Stoke's law on the left hand side of this equation to reduce the volume integral to the surface integral over ${\partial\Sigma_\infty}$ and ${\partial\Sigma_{+}}$ bounding the volume $\Sigma$
\begin{widetext}
    \begin{equation}
        \int_{\partial\Sigma_\infty} \left(K_{\mu;\beta}- {\cal B} g_{\mu\beta} \right) N^\mu {\hat T}^\beta \sqrt{|\gamma^{(\partial\Sigma_\infty)}|} d^2 y- \int_{\partial\Sigma_{+}} \left(K_{\mu;\beta}- {\cal B} g_{\mu\beta} \right) N^\mu {\hat T}^\beta \sqrt{|\gamma^{(\partial\Sigma_{+})}|} d^2 y = \int_\Sigma \left(R_{\nu\beta} K^\nu +3 {\cal B}_{,\beta} \right){\hat T}^\beta \sqrt{|\gamma^{(\Sigma)}|} d^3 x. \label{im43}
    \end{equation}
\end{widetext}
Here, $N^\mu$ is a unit spacelike vector normal to $\partial\Sigma$. Now, let us define the surface gravity, which reduces to the conformally invariant definition of Ref.~\cite{jacobson1993d} at the horizon
\begin{equation}
    \kappa= \left(K_{\mu;\beta}- 2 {\cal B} g_{\mu\beta}\right) N^\mu {\hat T}^\beta. \label{sg45}
\end{equation}

It is important to note that applying the definition of surface gravity to the linear Vaidya spacetime yields the result in Eq.~\eqref{gsg33}. On the non-null surfaces (like $\partial\Sigma_\infty$), the term containing the conformal factor $2 {\cal B}$ does not affect the surface gravity. Therefore, at the boundary $\partial\Sigma_\infty$ of Eq.~\eqref{im43}, the integrand within the integral can be interpreted as the surface gravity. Consequently, the integral over $\partial\Sigma_\infty$ reduces to the Komar expression for the total gravitating mass $M$ within the system~\cite{ep16}
\begin{multline}
    M = \frac{1}{4\pi} \int_\Sigma \left(R_{\nu\beta} K^\nu +3 {\cal B}_{,\beta} \right){\hat T}^\beta \sqrt{|\gamma^{(\Sigma)}|} d^3 x+ \\
    \frac{1}{4\pi} \int_{\partial\Sigma_{+}} \left(\kappa + {\cal B} g_{\mu\beta} N^\mu {\hat T}^\beta \right) \sqrt{|\gamma^{(\partial\Sigma_{+})}|} d^2 y. \label{tm46}
\end{multline}
This Komar mass calculated using the conformal Killing vector for dynamical spacetimes is equal to the conformal factor times the Komar mass of the conformal static spacetimes. Moreover, if we restrict ourselves to spherically symmetric spacetimes, we can write
\begin{multline}
    M = \frac{1}{4\pi} \int_\Sigma \left(R_{\nu\beta} K^\nu +3 {\cal B}_{,\beta} \right){\hat T}^\beta \sqrt{|\gamma^{(\Sigma)}|} d^3 x+\\
    \frac{\kappa}{4\pi} A- \frac{1}{4\pi} \int_{\partial\Sigma_{+}}  {\cal B} dA,
\end{multline}
where we have used the normalization condition $N_\mu {\hat T}^\mu= -1$ at the event horizon $\partial\Sigma_{+}$ and denoted its area by $A$. This formulae provides a generalization of the total mass for dynamical spacetimes that are conformal to static ones. We now make an assumption that black hole is slowly evaporating, such that we can neglect terms containing $\left(\partial M/\partial v\right)^2$ and $\partial^2 M/\partial v^2$. While restrictive, this approximation holds for macroscopic black holes where the evaporation rate is significantly smaller than the black hole mass itself~\cite{pf1,Dahal:2023suw}. This approximation implies a constant ${\cal B}$ (see Appendix~\ref{app5}), leading to a total mass formula identical in form to that of stationary spacetimes~\cite{Bardeen:1973}
\begin{equation}
    M = \frac{1}{4\pi} \int_\Sigma R_{\nu\beta} K^\nu {\hat T}^\beta \sqrt{|\gamma^{(\Sigma)}|} d^3 x+ \frac{\kappa- {\cal B}}{4\pi} A. \label{fl49}
\end{equation}
Here, $(\kappa- {\cal B})/4\pi$ represents the effective temperature, which incorporates contributions from both the semiclassical Hawking radiation and the classical black hole emission/accretion. This point will be discussed more in the next subsection.

\subsection{Differential first law} \label{s2a}

To observe if the variables appearing in the equation~\eqref{fl49} for the total mass behaves thermodynamically, we consider changes arising from parametric differences between infinitesimal diffeomorphic solutions
\begin{equation}
    g'_{\mu\nu}= g_{\mu\nu}+ h_{\mu\nu}.
\end{equation}
We can use the freedom in the choice of coordinates  of these conformally related solutions, such that the event horizon and conformal Killing vectors remain unchanged
\begin{equation}
    \delta K^\mu=0, \quad \delta K_\mu= h_{\mu\nu} K^\nu, \quad h_{\mu\nu}= - g_{\mu\alpha} g_{\nu\beta} \delta g^{\alpha\beta}.
\end{equation}
Not to restrict ourselves to spherically symmetric diffeomorphisms, we begin with Eq.~\eqref{tm46} with ${\cal B}$ set to constant (see above Eq.~\eqref{fl49}) and use the Einstein equations to obtain
\begin{multline}
    M = \int_\Sigma \left( 2 T_{\nu\beta}+ \frac{1}{8\pi} g_{\nu\beta} R^\mu_\mu \right) K^\nu {\hat T}^\beta \sqrt{|\gamma^{(\Sigma)}|} d^3 x+\\
    \frac{1}{4\pi} \int_{\partial\Sigma_{+}} \left(\kappa- {\cal B}\right) dA. \label{tm50}
\end{multline}
Here, we disregard the variation of the matter term $T_{\nu\beta}$ as it corresponds to the ordinary thermodynamics associated with matter. Specifically, considering matter as perfect fluids, its variation was worked out in Ref.~\cite{Bardeen:1973}. We thus begin by considering the variation of the term involving the scalar curvature $R^\mu_\mu$ (see Refs.~\cite{Bardeen:1973,wb13} for the derivation)
\begin{multline}
    \delta \left( g_{\nu\beta} R^\mu_\mu \sqrt{|\gamma^{(\Sigma)}|} K^\nu {\hat T}^\beta \right)=\\
    -\left( \left(R_{\mu\nu}- \frac{1}{2} g_{\mu\nu} R^\alpha_\alpha \right) h^{\mu\nu}+ 2 {h^\mu_{~~[\mu;\nu]}}^{;\nu} \right) K^\beta {\hat T}_\beta \sqrt{|\gamma^{(\Sigma)}|}. \label{vric51}
\end{multline}
Next, we expand the second term on the right hand side of this equation as
\begin{multline}
    K^\beta {h^\mu_{~~[\mu;\nu]}}^{;\nu}= {\mathcal L_K} \left(h_\mu^{~~[\mu;\beta]}\right)+ K^\nu_{~~;\nu} h_\mu^{~~[\mu;\beta]}+ \\
    \left(K^\beta h_\mu^{~~[\mu;\nu]}- K^\nu h_\mu^{~~[\mu;\beta]}\right)_{;\nu},
\end{multline}
where ${\mathcal L_K}$ is the Lie derivative along the vector $K^\mu$. One can obtain this identity simply by expanding the right hand side. Substituting this expansion and employing the Einstein equations, Eq.~\eqref{vric51} becomes
\begin{widetext}
    \begin{multline}
    \delta \left( g_{\nu\beta} R^\mu_\mu \sqrt{|\gamma^{(\Sigma)}|} K^\nu {\hat T}^\beta \right)=\\
    -\left( T_{\mu\nu} h^{\mu\nu} K^\beta+ 2 {\mathcal L_K} \left(h_\mu^{~~[\mu;\beta]}\right)+ 2 K^\nu_{~~;\nu} h_\mu^{~~[\mu;\beta]}+ 2 \left(K^\beta h_\mu^{~~[\mu;\nu]}- K^\nu h_\mu^{~~[\mu;\beta]}\right)_{;\nu} \right) {\hat T}_\beta \sqrt{|\gamma^{(\Sigma)}|}. \label{vricci53}
\end{multline}
\end{widetext}
Just for the same reason mentioned below Eq.~\eqref{tm50}, we drop the first term on the right hand side of this equation. This term, along with the first term on the right hand side of Eq.~\eqref{tm50}, contributes solely to the ordinary matter thermodynamics (see Ref.~\cite{Bardeen:1973}). The simplification of the second and third terms is presented in Appendix~\ref{app6}. The final term exhibits antisymmetry and thus can be converted to a surface integral by the application of Stoke's theorem
\begin{widetext}
    \begin{multline}
    -\frac{1}{4\pi} \int_\Sigma \left(K^\beta h_\mu^{~~[\mu;\nu]}- K^\nu h_\mu^{~~[\mu;\beta]}\right)_{;\nu} {\hat T}_\beta \sqrt{|\gamma^{(\Sigma)}|} d^3 x=\\
    - \frac{1}{4\pi} \int_{\partial\Sigma_\infty} \left(K^\beta h_\mu^{~~[\mu;\nu]}- K^\nu h_\mu^{~~[\mu;\beta]}\right) N_\nu {\hat T}_\beta \sqrt{|\gamma^{(\partial\Sigma_\infty)}|} d^2 y+ \frac{1}{4\pi} \int_{\partial\Sigma_{+}} \left(K^\beta h_\mu^{~~[\mu;\nu]}- K^\nu h_\mu^{~~[\mu;\beta]}\right) N_\nu {\hat T}_\beta \sqrt{|\gamma^{(\partial\Sigma_{+})}|} d^2 y, \label{vsi54}
\end{multline}
where ${\partial\Sigma_\infty}$ and ${\partial\Sigma_{+}}$ denote the surfaces enclosing the volume $\Sigma$. The integral over $\partial\Sigma_\infty$ contributes to the variation in mass~\cite{Wald:1984rg} (for dynamical spacetimes considered here, we can easily verify this relation using Eq.~\eqref{dvf1})
\begin{equation}
    - \frac{1}{4\pi} \int_{\partial\Sigma_\infty} \left(K^\beta h_\mu^{~~[\mu;\nu]}- K^\nu h_\mu^{~~[\mu;\beta]}\right) N_\nu {\hat T}_\beta \sqrt{|\gamma^{(\partial\Sigma_\infty)}|} d^2 y= -\delta M. \label{si55}
\end{equation}
We will show in Appendix~\ref{app6} that the integral over the horizon $\partial\Sigma_{+}$ yields
\begin{equation}
\frac{1}{4\pi} \int_{\partial\Sigma_{+}} \left(K^\beta h_\mu^{~~[\mu;\nu]}- K^\nu h_\mu^{~~[\mu;\beta]}\right) N_\nu {\hat T}_\beta \sqrt{|\gamma^{(\partial\Sigma_{+})}|} d^2 y= \frac{1}{4\pi} \int_{\partial\Sigma_{+}} \left( 4 {\cal B} k_0- 2 {\cal B} (k_3+ k_6) - \delta\kappa + {\cal B} h_{\beta\gamma} {\hat T}^\gamma N^\beta \right) dA. \label{si56}
\end{equation}
Incorporating results from Eqs.~\eqref{vsi54}, \eqref{si55}, \eqref{si56} and \eqref{lf15} and substituting them back into Eq.~\eqref{vricci53}, we obtain
\begin{multline}
    \frac{1}{8\pi} \int_\Sigma \delta \left( g_{\nu\beta} R^\mu_\mu \sqrt{|\gamma^{(\Sigma)}|} K^\nu {\hat T}^\beta \right) d^3 x=\\
    -\delta M + \frac{1}{4\pi} \int_{\partial\Sigma_{+}} \left( 4 {\cal B} k_0- 2 {\cal B} (k_3+ k_6) - \delta\kappa + {\cal B} h_{\beta\gamma} {\hat T}^\gamma N^\beta \right) dA +\frac{1}{4\pi} \int_{\partial\Sigma_{+}} 3 {\cal B} \left(k_1+ k_2\right) dA,
\end{multline}
where the symbols ($k_i$'s) used here are introduced in Appendix~\ref{app62}. Substituting this expression into Eq.~\eqref{tm50} then yields the variation in total mass
\begin{equation}
    \delta M= \frac{1}{8\pi} \int_{\partial\Sigma_{+}} 3 {\cal B} \left(k_1+ k_2\right) dA+ \frac{1}{8\pi} \int_{\partial\Sigma_{+}} \left( 4 {\cal B} k_0- 2 {\cal B} (k_3+ k_6) +{\cal B} h_{\beta\gamma} {\hat T}^\gamma N^\beta \right) dA+ \frac{1}{8\pi} \int_{\partial\Sigma_{+}} \left(\kappa- {\cal B}\right) \delta(dA)+ \textrm{matter terms}.
\end{equation}
\end{widetext}
Furthermore, substituting the relation $k_1+ k_2= -2 h^{\mu\nu} {\hat T}_\mu N_\nu \eqdef 2 k_0$ derived in Eq.~\eqref{dv61} gives the final expression for the differential mass of dynamical spacetimes conformal to static ones
\begin{multline}
    \delta M= \frac{1}{8\pi} \int_{\partial\Sigma_{+}} \left(9 k_0- 2 (k_3+ k_6)\right) {\cal B} dA+ \\
    \frac{1}{8\pi} \int_{\partial\Sigma_{+}} \left(\kappa- {\cal B}\right) \delta(dA)+ \textrm{matter terms}. \label{t57}
\end{multline}
In contrast to the static case, where the pure evaporation by Hawking effects leads to a constant temperature of $\kappa$, dynamically radiating black holes has an effective temperature of $\kappa-{\cal B}$. There is also a notion of surface gravity defined in the literature, which is associated with this effective temperature~\cite{jacobson1993d}. Consider the example of Vaidya black holes, where, besides Hawking radiation, continuous emission/accretion of null fluid contributes to mass variation. This accretion/evaporation modifies the mass by $\int_{\partial\Sigma_{+}} \Phi dA$, where $\Phi$ represents the energy emitted per unit area. Consequently, the differential mass takes the form
\begin{equation}
    \int_{\partial\Sigma_{+}} \Phi \delta(dA).
\end{equation}
Here, $\Phi\propto -{\cal B}$ and $\delta\Phi$ vanish because the energy emitted per unit area is constant.
This naturally explains the presence of the effective temperature in the differential mass formula for dynamical black holes. However, the appearance of the first term was not expected. This term, absent for static black holes, captures the mass change due to the black hole's (classical) dynamics. To explain this, we can draw an analogy: in ultra-stationary spacetimes, gravitational field exhibits perfect resemblance to a dielectric medium~\cite{hereirdo:2006}. However, this exact correspondence breaks down in general. Though some aspects remain reminiscent of a dielectric medium, the full nature of spacetimes become significantly more intricate. Similarly, while the horizon behaves exactly thermodynamically in the stationary limit, its true nature might be more complex.

\section{Discussion} \label{discus}

We have studied the problem of particle production in dynamical spacetimes through explicit calculations of Bogoliubov coefficients. We have also calculated the total mass of a dynamical black hole and its differential variation to observe whether its mechanics transcends the laws of black hole thermodynamics. To the best of our knowledge, this work provides the first exact derivation of Hawking radiation and calculation of total energy rooted in first principles, in dynamical spacetimes. As mentioned above, these tasks are crucial for addressing conceptual issues and exploring new frontiers in black hole physics. One advantage, as Vaidya metric itself is a candidate solution to the semiclassical backreaction problem in the leading order approximation, using this metric inherently incorporates backreaction effects into the derivation of Hawking radiation, and makes the calculations inherently self-consistent.

Our calculation confirms that the surface where the retarded null coordinate is infinitely large (because of which an approximate e-folding relation holds between the retarded coordinate $u$ and the Kruskal coordinate $U$) is necessary to describe the Hawking effect~\cite{blsm6}. This surface closely approximates the event horizon. Moreover, formulation of the laws of black hole mechanics here shows that it is the event horizon that behaves thermodynamically. We have also confirmed that black hole emits Hawking radiation analogous to a gray body of temperature proportional to the geometric surface gravity. Consideration of the classical black hole dynamics causes observer to measure different temperatures, some of which correspond to different notions of surface gravity appearing in the literature (see Eqs.~\eqref{t35} and \eqref{t57}). Also, in spacetimes that are conformal transformations of static ones, there exists a frame of reference corresponding to the conformal $(T,R)$ coordinates (see Eq~\eqref{cs3}). This reference frame allow us to formulate the laws of thermodynamics akin to those for static black holes, without relying on the adiabaticity assumption.

We have also discussed the possibility that dynamical spacetimes, even the spherically symmetric ones exhibit the possibility of superradiant scattering. Detailed investigation of this phenomenon will be reserved for future studies. While our study employs dynamical spacetimes to investigate particle production, the framework falls well within the regime where geometric optics approximation is considered valid in the literature~\cite{fabbri:2005,harlow2014}. This work does not explicitly address its validity here. It offers neither further insights into the application of this approximation, nor into the trans-Planckian problem~\cite{Jacobson:2005,Helfer:2003} that arises because of infinite redshift at the horizon. Our approach is applicable to spacetimes which are conformal to the static ones and which possess a null surface whose retarded coordinate is infinitely large. However, for horizonless compact objects, we may need to consider matter source contributions only, necessitating a full statistical treatment to understand their thermodynamics.~\cite{aydemir:2023}.

\acknowledgements

We thank Ioannis Soranidis for his useful comments. PKD and SM are supported by an International Macquarie University Research Excellence Scholarship.

\section*{Data Availability Statement}

Data sharing is not applicable to this article as no new data were created or analyzed in this study.

\appendix

\begin{widetext}

\section{Evaporating Vaidya black hole in static coordinates} \label{app1}

Take the metric of Eq.~\eqref{v1} and apply following coordinate transformation relationships
\begin{equation}\begin{aligned} \label{r3}
\alpha v=& r_0 - r_0 e^{-\alpha {\cal T}/r_0}, \qquad dv= e^{-\alpha {\cal T}/r_0} d{\cal T},\\
    r=& R e^{-\alpha {\cal T}/r_0}, \qquad dr= e^{-\alpha {\cal T}/r_0}\left(dR- \frac{\alpha R}{r_0} d{\cal T}\right).
\end{aligned}\end{equation}
We then obtain
\begin{equation}\begin{aligned}
    ds^2=& e^{-2\alpha {\cal T}/r_0} \left(-\left(1-\frac{r_0}{R}+ \frac{2\alpha R}{r_0}\right) d{\cal T}^2+ 2 d{\cal T} dR+ R^2 d\Omega^2\right)\\
        =& e^{-2\alpha {\cal T}/r_0} \left(-\left(1-\frac{r_0}{R}+ \frac{2\alpha R}{r_0}\right) dT^2+ \frac{1}{1-\frac{r_0}{R}+ \frac{2\alpha R}{r_0}} dR^2+ R^2 d\Omega^2\right),
\end{aligned}\end{equation}
where we have defined the Schwarzschild coordinate
\begin{equation}
    dT= d{\cal T}- \frac{1}{1-\frac{r_0}{R}+ \frac{2\alpha R}{r_0}} dR, \label{ac6}
\end{equation}
which is perfectly integrable. Furthermore, let us define the retarded coordinates
\begin{equation}
    du= d{\cal T}- \frac{2}{1-\frac{r_0}{R}+ \frac{2\alpha R}{r_0}} dR \equiv dT- \frac{1}{1-\frac{r_0}{R}+ \frac{2\alpha R}{r_0}} dR, \label{app3}
\end{equation}
such that, the metric in this coordinate takes the form
\begin{equation}
    ds^2= e^{-2\alpha {\cal T}/r_0} \left(-\left(1-\frac{r_0}{R}+ \frac{2\alpha R}{r_0}\right) d{\cal T} du+ R^2 d\Omega^2\right).
\end{equation}
This is the retarded coordinate for both the Vaidya metric and its conformal counterpart.

\section{QFT in linear Vaidya}

\subsection{Solutions of radial differential equation} \label{appde2}

We start with the radial differential equation~\eqref{de14}, expressed in terms of tortoise coordinate. Near the event horizon of the static metric, $1-r_0/R+ 2\alpha R/r_0\to 0$, this equation simplifies to the form
\begin{equation}
    \frac{d^2{\tilde\phi}_R}{d{R^*}^2}+ k^2= 0,
\end{equation}
and the corresponding solution is an elementary function
\begin{equation}
    {\tilde\phi}_R= e^{\pm i k R^*}. \label{eq15}
\end{equation}
So, let us write the general solution of Eq.~\eqref{de14} as
\begin{equation}
    {\tilde\phi}^\pm_R= e^{\pm i k R^*} {\cal G}^\pm(R).
\end{equation}
Substituting this ansatz back into the original equation~\eqref{de14} gives the differential equation for $\cal G(R)$. Let us first consider the case with the positive sign on Eq.~\eqref{eq15}
\begin{equation}
    \frac{d^2 {\cal G}^+(R)}{d{R^*}^2}+ \left(2 i k+ \frac{2}{R} \left(1-\frac{r_0}{R}+ \frac{2\alpha R}{r_0} \right) \right) \frac{d{\cal G}^+(R)}{d{R^*}}+ \left(1-\frac{r_0}{R}+ \frac{2\alpha R}{r_0} \right) \left(\frac{2 i k}{R}- \frac{l(l+1)}{R^2}\right){\cal G}^+(R) =0.
\end{equation}
Reverting back to the original radial coordinate $R$, we obtain
\begin{equation}
    \frac{d^2 {\cal G}^+}{dR^2}+ \frac{2 R r_0- r_0^2+ (6\alpha+ 2 i k r_0) R^2}{R(r_0 R- r_0^2+ 2\alpha R^2)} \frac{d{\cal G}^+}{dR}+ \frac{2 i k R- l(l+1)}{R^2 \left(1-r_0/R+ 2\alpha R/r_0\right)} {\cal G}^+ = 0.
\end{equation}
Let us rewrite this equation in the form
\begin{equation}
    \frac{d^2 {\cal G}^+}{dR^2}+ \left( \frac{1}{R}+ \frac{1- 2 i k r_- \big/\sqrt{1+ 8\alpha}}{ R- r_-}+ \frac{1+ 2 i k r_+ \big/\sqrt{1+ 8\alpha}}{R- r_+}\right) \frac{d{\cal G}^+}{dR}
    + \frac{r_0}{2\alpha R} \frac{2 i k R- l(l+1)}{\left(R- r_-\right) \left(R- r_+\right)} {\cal G}^+ = 0.
\end{equation}
Rescaling the variable $R \to R r_+$ and $k \to k/r_+$, this equation reduces to the form
\begin{equation}
    \frac{d^2 {\cal G}^+}{dR^2}+ \left( \frac{1}{R}+ \frac{1- 2 i k r_- \big/(r_+\sqrt{1+8\alpha} ) }{R- r_- \big/r_+}+ \frac{1+ 2 i k \big/\sqrt{1+8\alpha}}{R-1}\right) \frac{d{\cal G}^+}{dR}
    + \frac{2 i k R- l(l+1)}{R \left(R- r_- \big/r_+ \right) \left(R-1\right)} \frac{2}{\sqrt{1+8\alpha}-1} {\cal G}^+ = 0. \label{b7}
\end{equation}
We thus have the Heun differential equation of the form
\begin{equation}
    \frac{d^2 {\cal G}^+}{dR^2}+ \left( \frac{\gamma}{R}+ \frac{\epsilon}{ R- r_- \big/r_+}+ \frac{\delta}{R-1}\right) \frac{d{\cal G}^+}{dR}\\
    + \frac{\zeta\beta R- q}{R \left(R- r_- \big/r_+ \right) \left(R-1\right)} {\cal G}^+ = 0.
\end{equation}
Comparing the coefficients, we find
\begin{equation}\begin{aligned}
    &\gamma=1, \quad \epsilon= 1- \frac{2 i k}{\sqrt{1+8\alpha}} \frac{r_-}{r_+}, \quad \delta= 1+ \frac{2 i k}{\sqrt{1+8\alpha}},\\
    &\zeta=1+ \frac{i k r_0}{2 \alpha r_+}+ \sqrt{1-\frac{k^2 r_0^2}{4 \alpha^2 r_+^2}}, \quad \beta=1+ \frac{i k r_0}{2 \alpha r_+}- \sqrt{1-\frac{k^2 r_0^2}{4 \alpha^2 r_+^2}}, \quad q= \frac{l(l+1) r_0}{2 \alpha r_+}.
\end{aligned}\end{equation}
Heun differential equation contains four regular singular points
\begin{equation}
    R=0, \quad R=1, \quad R= \infty, \quad R= \frac{r_-}{r_+}.
\end{equation}
Near each of these points, Heun differential equation has the power series solutions with two different exponents. For example, near the horizon (where the exponents are $0$ and $1-\delta$), we can write the solution as
\begin{equation}\begin{aligned}
    &{\cal G}_{11}^+= HG\left(\frac{\sqrt{1+8\alpha} r_0}{2 \alpha r_+},\zeta\beta-q;\zeta,\beta,\delta,\gamma;1-R\right),\\
    &{\cal G}_{12}^+= (1-R)^{1-\delta} HG\left( \frac{\sqrt{1+8\alpha} r_0}{2 \alpha r_+},\left(\frac{\sqrt{1+8\alpha} r_0}{2 \alpha r_+} \gamma +\epsilon\right)(1-\delta)+ \zeta\beta-q;\zeta+1-\delta,\beta+1-\delta,2-\delta,\gamma;1-R\right).
\end{aligned}\end{equation}
Let us again consider the equation with the negative sign in Eq.~\eqref{eq15} and write the differential equation as

Again, we substitute the ansatz of Eq.~\eqref{eq15} (but this time taking its negative sign) into the original equation~\eqref{de14}, to obtain
\begin{equation}
    \frac{d^2 {\cal G}^-(R)}{d{R^*}^2}+ \left( -2 i k+ \frac{2}{R} \left(1-\frac{r_0}{R}+ \frac{2\alpha R}{r_0} \right) \right) \frac{d{\cal G}^-(R)}{d{R^*}}- \left(1-\frac{r_0}{R}+ \frac{2\alpha R}{r_0} \right) \left(\frac{2 i k}{R}+ \frac{l(l+1)}{R^2}\right){\cal G}^-(R) =0.
\end{equation}
Reverting back to the original radial coordinate $R$, we obtain
\begin{equation}
    \frac{d^2 {\cal G}^-}{dR^2}+ \frac{2 R r_0- r_0^2+ (6\alpha- 2 i k r_0) R^2}{R(r_0 R- r_0^2+ 2\alpha R^2)} \frac{d{\cal G}^-}{dR}- \frac{2 i k R+ l(l+1)}{R^2 \left(1-r_0/R+ 2\alpha R/r_0\right)} {\cal G}^- = 0.
\end{equation}
Let us rewrite this equation in the form
\begin{equation}
    \frac{d^2 {\cal G}^-}{dR^2}+ \left( \frac{1}{R}+ \frac{1+ 2 i k r_- \big/\sqrt{1+ 8\alpha}}{ R- r_-}+ \frac{1- 2 i k r_+ \big/\sqrt{1+ 8\alpha}}{R- r_+}\right) \frac{d{\cal G}^-}{dR}
    - \frac{r_0}{2\alpha R} \frac{2 i k R+ l(l+1)}{\left(R- r_-\right) \left(R- r_+\right)} {\cal G}^- = 0.
\end{equation}
Rescaling the variable $R \to R r_+$ and $k \to k/r_+$, this equation reduces to the form
\begin{equation}
    \frac{d^2 {\cal G}^-}{dR^2}+ \left( \frac{1}{R}+ \frac{1+ 2 i k r_- \big/(r_+\sqrt{1+8\alpha} ) }{R- r_- \big/r_+}+ \frac{1- 2 i k \big/\sqrt{1+8\alpha}}{R-1}\right) \frac{d{\cal G}^-}{dR}
    - \frac{2 i k R+ l(l+1)}{R \left(R- r_- \big/r_+ \right) \left(R-1\right)} \frac{2}{\sqrt{1+8\alpha}-1} {\cal G}^- = 0.
\end{equation}
This is the Heun differential equation, which is the complex conjugate of the corresponding equation~\eqref{b7}. Let us denote its solutions as ${\cal G}_{11}^-$ and ${\cal G}_{12}^-$.

\subsection{Normalization of asymptotic mode functions} \label{appm4}

Asymptotic modes near null infinities are given in equations ~\eqref{im20} and ~\eqref{im22}. These incoming and outgoing modes with different $k,l,m$ are orthogonal to each other and then, we can obtain the normalization factors for each of the modes. The scalar product $(f_{I},f_{J})$, where $I=\{\text{in/out},k\}$, is defined as 
\begin{align}
    (f_{I},f_{J})=-i\int_{\Sigma} d^3 x \sqrt{-g} \,\,g^{0\mu} \,\,f_{I}^*(t,\mathbf{x}) \overleftrightarrow{\partial_\mu} f_{J}(t,\mathbf{x}),
\end{align}
where the integration is performed on the constant time hypersurface $\Sigma$ and
\begin{align}
    f_{I}^*(t,\mathbf{x}) \overleftrightarrow{\partial_\mu} f_{J}(t,\mathbf{x})\equiv f_{I}^*(t,\mathbf{x}) \partial_\mu f_{J}(t,\mathbf{x}) - f_{J}(t,\mathbf{x}) \partial_\mu f_{I}^*(t,\mathbf{x}).
\end{align}
Here $\mathbf{x}$ denotes the three coordinates on the hypersurface $\Sigma$.

We will be performing integral of the scalar product on the spacelike hypersurface $T=\text{const.}$, which is given as 
\begin{equation}
    \begin{aligned}
        (f^{\text{in}}_{k l m},f^{\text{in}}_{k' l' m'}) &=-i\int_\Sigma R^2 \sin \theta \,\,dR\,\, d\theta\,\, d\phi\,\, e^{\frac{4 \alpha (T+R_*)}{r_0}} g^{TT} f_{klm}^*\overleftrightarrow{\partial_T} f_{k'l'm'} \\
    &=\int_\Sigma \left[\sin \theta \,d\theta \,d\phi \,Y_{lm}(\theta,\phi) \,Y_{l'm'}(\theta,\phi)\right]\,\left[dR_* \,e^{i(k-k')R_*}\right]\, e^{i(k-k')T} C_{klm}\,C_{k'l'm'}^*\,(k+k')\\
    &=4\pi k \,\vert C_{l m}(k)\vert^2 \,\,\delta_{l l'}\,\,\delta_{m m'}\,\,\delta(k-k') ,\label{eqn:normcondn1}
    \end{aligned}
\end{equation}
where
\begin{equation}
    \int_{\Sigma} = \int_{R= 2 r_0/(1+\sqrt{1+8\alpha})}^{\infty} \int_{\theta=0}^{\pi/2} \int_{\phi=0}^{2\pi} 
\end{equation}
Thus, the normalized asymptotic modes near past null infinity take the form
\begin{align}
    f^{\text{in}}_{klm}=\frac{1}{\sqrt{4\pi |k|}\, R} \, e^{-ik (T+R_*)} \,e^{\frac{\alpha(T+R_*)}{r_0}}\,Y_{lm}(\theta,\phi) .\label{eqn:basissoln}
\end{align}

Similarly, we can show that 
\begin{align}
    (f^{\text{in}}_{k l m},f^{\text{out}}_{k' l' m'})=0 \,\, , \,\,(f^{\text{out}}_{k l m},f^{\text{out}}_{k' l' m'}) = 4\pi k \,\vert D_{l m}(k)\vert^2 \,\,\delta_{l l'}\,\,\delta_{m m'}\,\,\delta(k-k') .\label{eqn:normcondn2}
\end{align}
Thus, the outgoing modes near null infinity can be normalized from the second part of equation (\ref{eqn:normcondn2}) to give
\begin{align}
    f^{\text{out}}_{klm}=\frac{1}{\sqrt{4\pi |k|}\, R} \, e^{-ik (T-R_*)} \,e^{\frac{\alpha(T+R_*)}{r_0}}\,Y_{lm}(\theta,\phi). \label{eqn:basissoln2}
\end{align}

If the basis states $f^{\text{in}}_{klm}$ and $f^{\text{out}}_{klm}$ are complete, then the identity operator should be given as
\begin{align}
    I=&\int_{0}^{\infty} dk \sum_{l,m}f^{\text{in}}_{lm}(k)\left( f^{\text{in}}_{lm}(k),\cdot\right)-\int_{-\infty}^{0} dk \sum_{l,m}f^{\text{in}}_{lm}(k)\left( f^{\text{in}}_{lm}(k),\cdot\right) \notag\\
    &+\int_{0}^{\infty} dk \sum_{l,m}f^{\text{out}}_{lm}(k)\left( f^{\text{out}}_{lm}(k),\cdot\right)-\int_{-\infty}^{0} dk \sum_{l,m}f^{\text{out}}_{lm}(k)\left( f^{\text{out}}_{lm}(k),\cdot\right) .
\end{align}
Applying the identity operator $I$ on an arbitrary function $\Phi(x)$, we obtain
\begin{align}
    \Phi(x)=&\int_{0}^{\infty} dk \sum_{l,m}\left[f^{\text{in}}_{klm}(x)\left( f^{\text{in}}_{klm},\Phi\right)-f^{\text{in}}_{(-k)lm}(x)\left( f^{\text{in}}_{(-k)lm},\Phi\right)\right.\notag\\
    &\,\,\,\,\,\,\,\,\,\,\,\,\,\,\,\,\,\,\,\,\,\,\,\,\,\,\,\,\,\,\,\, \left.+ f^{\text{out}}_{klm}(x)\left( f^{\text{out}}_{klm}(k),\Phi\right)-f^{\text{out}}_{(-k)lm}(x)\left( f^{\text{out}}_{(-k)lm},\Phi\right)\right] .
\end{align}
If the states $f_{klm}$ are complete, then the above equation implies that the following relations should be satisfied~\cite{Boca:2011}
\begin{align}
    & \int_{0}^{\infty} dk \sum_{l,m}\left[f^{\text{in}}_{klm}(x)f^{*\text{in}}_{klm}(x')-f^{\text{in}}_{(-k)lm}(x)f^{*\text{in}}_{(-k)lm}(x')+f^{\text{out}}_{klm}(x)f^{*\text{out}}_{klm}(x')-f^{\text{out}}_{(-k)lm}(x)f^{*\text{out}}_{(-k)lm}(x')\right]=0, \label{eqn:compcond1}\\
    & \int_{0}^{\infty} dk \sum_{l,m}\left[f^{\text{in}}_{klm}(x)\partial_T f^{*\text{in}}_{klm}(x')-f^{\text{in}}_{(-k)lm}(x)\partial_T f^{*\text{in}}_{(-k)lm}(x')+f^{\text{out}}_{klm}(x)\partial_T f^{*\text{out}}_{klm}(x')-f^{\text{out}}_{(-k)lm}(x)\partial_T f^{*\text{out}}_{(-k)lm}(x')\right] \notag\\
    & \,\,\,\,\,\,\,\,\,\,\,\,\,\,\,\,\,\,\,\,\,\,\,\,\,\,\,\,\,\,\,\,\,\,\,\,\,\,\,\,\,\,\,\,\,\,\,\,=i\,\frac{e^{\frac{2\alpha}{r_0}(T+R_*)}}{R^2}\left(1-\frac{r_0}{R}+\frac{2\alpha R}{r_0}\right)\delta(R-R')\frac{1}{\sin \theta}\delta(\theta-\theta')\delta(\phi-\phi')  .\label{eqn:compcond2}
\end{align}
Equation~\eqref{eqn:compcond1} can be readily verified by substituting the expressions for $f^\text{in}_{klm}$ and $f^\text{out}_{klm}$ respectively obtained in equations (\ref{eqn:basissoln}) and (\ref{eqn:basissoln2}). The contribution from the incoming and outgoing modes cancel each other and thus it is equal to $0$. Performing the same exercise to verify equation~\eqref{eqn:compcond2}, we obtain
\begin{align}
    &\int_{0}^{\infty} dk \sum_{l,m}\left[f^{\text{in}}_{klm}(x)\partial_T f^{*\text{in}}_{klm}(x')-f^{\text{in}}_{(-k)lm}(x)\partial_T f^{*\text{in}}_{(-k)lm}(x')+f^{\text{out}}_{klm}(x)\partial_T f^{*\text{out}}_{klm}(x')-f^{\text{out}}_{(-k)lm}(x)\partial_T f^{*\text{out}}_{(-k)lm}(x')\right] \notag\\
    &=\frac{i}{\pi R R'}e^{\frac{\alpha}{r_0}(2T+R_*+R_*')} \int_{0}^{+\infty} dk \left(\cos(k(R_*'-R_* ))\right)\delta(\Omega-\Omega')\label{eqn:compcond2lastint}\\
    &=\frac{i}{R^2}e^{\frac{2\alpha}{r_0}(T+R_*)}\left(1-\frac{r_0}{R}+\frac{2\alpha R}{r_0}\right)\delta(R-R')\frac{1}{\sin \theta}\delta(\theta-\theta')\delta(\phi-\phi') ,
\end{align}
where the integral in equation (\ref{eqn:compcond2lastint}) can be computed using the identity $\int_{0}^{\infty} dt\, \cos(t x) = \pi \delta(x)$.

\subsection{Tracing null ray back from future to past null infinity} \label{app2}

Radial null geodesics for the static metric in $(T,R)$ coordinates, given in Eq.~\eqref{cs3}, is given by
\begin{equation}
    \frac{dR}{d\lambda}= \pm E,
\end{equation}
where positive/negative sign corresponds to the outgoing/incoming geodesics. Along the incoming radial null geodesics defined by ${\cal T}= {\cal T}_1$, we can calculate the retarded null coordinate using Eq.~\eqref{app3}
\begin{equation}
    \frac{du}{d\lambda}= \frac{dT}{d\lambda}- \frac{dR^*}{d\lambda}, \quad \frac{dR^*}{d\lambda}= -\frac{E}{1-\frac{r_0}{R}+ \frac{2\alpha R}{r_0}}.
\end{equation}
Substituting the value of $dT/d\lambda$ and simplification gives
\begin{equation}
    \frac{du}{d\lambda}= \frac{2 E}{1-\frac{r_0}{R}+ \frac{2\alpha R}{r_0}}. \label{d3}
\end{equation}
As, $dR/d\lambda= -E$ along this geodesics, we have
\begin{equation}
    R- \frac{-1+\sqrt{1+8\alpha}}{4\alpha} r_0= -E \lambda.
\end{equation}
Here, we have chosen the value of the parameter $\lambda$, such that, it is zero at the event horizon and negative outside. Hence,
\begin{equation}
    \frac{1}{1-\frac{r_0}{R}+ \frac{2\alpha R}{r_0}}= \frac{R r_0}{2\alpha \left(R- r_-\right) \left(R- r_+\right)}= \frac{-E \lambda+ r_+}{2\alpha \left(-E\lambda+ \frac{\sqrt{1+8\alpha}}{2\alpha} r_0\right) (-E\lambda)} r_0.
\end{equation}
Substituting this relation back into Eq.~\eqref{d3} gives
\begin{equation}
    \frac{du}{d\lambda}= \frac{-E \lambda+ r_+}{\alpha \left(-E\lambda+ \frac{\sqrt{1+8\alpha}}{2\alpha} r_0\right) (-\lambda)} r_0.
\end{equation}
Integrating this quantity and then expanding near the horizon, we obtain
\begin{equation}
    u \approx \frac{1- \sqrt{1+ 8\alpha}}{2\alpha \sqrt{1+ 8\alpha}} r_0 \ln\left(\lambda/K_1\right), \label{d7}
\end{equation}
for some constant $K_1$. Because ${\cal I}^-$ is far from the collapsing body, the coordinate ${\cal T}$ is itself an affine parameter along ${\cal I}^-$. Therefore, ${\cal T}- {\cal T}_0$ must be related to the affine separation $\lambda$ between $u({\cal T})- u({\cal T}_0)$ on ${\cal I}^+$ by
\begin{equation}
    {\cal T}_0- {\cal T}= K_2 \lambda,
\end{equation}
for some constant $K_2$. Substituting the value of $\lambda$ back into Eq.~\eqref{d7}, we get
\begin{equation}
    u({\cal T})= \frac{1- \sqrt{1+ 8\alpha}}{2\alpha \sqrt{1+ 8\alpha}} r_0 \ln\left(\frac{{\cal T}_0- {\cal T}}{K}\right), \label{a13}
\end{equation}
where $K= K_1 K_2$ is a constant characterizing the affine parametrization of the geodesics when it is near ${\cal I}^+$ and ${\cal{I}}^-$.

Now, we substitute this value of the phase into Eqs.~\eqref{bc26} and \eqref{bc27} to simplify them further
\begin{equation}\begin{aligned}
    \alpha_{kk'}=& C \int_{-\infty}^{{\cal T}_0} d{\cal T} \left(\frac{k'}{k}\right)^{1/2} e^{2 i k r_+ \big/ \sqrt{1+ 8\alpha} \ln \left(\frac{{\cal T}_0-{\cal T}}{K}\right)} e^{i k' {\cal T}} ,\\
    \beta_{kk'}=& C \int_{-\infty}^{{\cal T}_0} d{\cal T} \left(\frac{k'}{k}\right)^{1/2} e^{2 i k r_+ \big/ \sqrt{1+ 8\alpha} \ln \left(\frac{{\cal T}_0-{\cal T}}{K}\right)} e^{-i k' {\cal T}} .
\end{aligned}\end{equation}
The subsequent simplification steps closely mirror those employed in the derivation of Hawking radiation in Schwarzschild spacetime. We change the variables $s= {\cal T}_0-{\cal T}$ in the first equation and $s= {\cal T}- {\cal T}_0$ in the second equation, to obtain
\begin{equation}\begin{aligned}
    \alpha_{kk'}=& -C \int_{\infty}^{0} ds \left(\frac{k'}{k}\right)^{1/2} e^{-i k' s} e^{i k' {\cal T}_0} e^{2 i k r_+ \big/ \sqrt{1+ 8\alpha} \ln \left(s/K\right)} ,\\
    \beta_{kk'}=& C \int_{-\infty}^{0} ds \left(\frac{k'}{k}\right)^{1/2} e^{-i k' s} e^{-i k' {\cal T}_0 } e^{2 i k r_+ \big/ \sqrt{1+ 8\alpha} \ln \left(-s/K \right)} .
\end{aligned}\end{equation}
As the coefficient $\alpha_{kk'}$ is the Fourier transform of the function that vanishes for $s'>0$, it is analytic on the lower half of the complex $k'$ plane. So, we can substitute $s= i s'$ in these integrals, to get (under this substitution, the integral from $0$ to $\infty$ is equal to the integral from $-i\infty$ to $0$)
\begin{equation}\begin{aligned}
    \alpha_{kk'}=& -i C \int_{-\infty}^{0} ds' \left(\frac{k'}{k}\right)^{1/2} e^{ k' s'} e^{i k' {\cal T}_0} e^{2 i k r_+ \big/ \sqrt{1+ 8\alpha} \ln \left(i s'/K\right)} ,\\
    \beta_{kk'}=& i C \int_{-\infty}^{0} ds' \left(\frac{k'}{k}\right)^{1/2} e^{ k' s'} e^{-i k' {\cal T}_0 } e^{2 i k r_+ \big/ \sqrt{1+ 8\alpha} \ln \left(-i s'/K\right)} .
\end{aligned}\end{equation}
$\alpha_{kk'}$ has a logarithmic branch point at $s'=0$. Taking the cut in the complex plane along the negative real axis to define a single values natural logarithm function, we find that for $s'<0$
\begin{equation}\begin{aligned}
    \ln\left(\frac{i s'}{K}\right)=& \ln\left(-\frac{i |s'|}{K}\right)= -\frac{i \pi}{2}+ \ln\left(\frac{|s'|}{K}\right),\\
    \ln\left(\frac{-i s'}{K}\right)=& \ln\left(\frac{i |s'|}{K}\right)= \frac{i \pi}{2}+ \ln\left(\frac{|s'|}{K}\right).
\end{aligned}\end{equation}
Substituting these results into above equations, we get
\begin{equation}\begin{aligned}
    \alpha_{kk'}=& -i C e^{i k' {\cal T}_0} e^{-\frac{\pi k (1- \sqrt{1+8\alpha}) r_0}{4\alpha \sqrt{1+8\alpha} }} \int_{-\infty}^{0} ds' \left(\frac{k'}{k}\right)^{1/2} e^{ k' s'}  e^{2 i k r_+ \big/ \sqrt{1+ 8\alpha} \ln \left(|s'|/K\right)} ,\\
    \beta_{kk'}=& i C e^{-i k' {\cal T}_0} e^{\frac{\pi k (1- \sqrt{1+8\alpha}) r_0}{4\alpha \sqrt{1+8\alpha} }} \int_{-\infty}^{0} ds' \left(\frac{k'}{k}\right)^{1/2} e^{ k' s'} e^{2 i k r_+ \big/ \sqrt{1+ 8\alpha} \ln \left(|s'|/K\right)} .
\end{aligned}\end{equation}
From these relations, Eq.~\eqref{bc28} follows.

\section{Slowly evolving spacetimes conformal to static ones} \label{app5}

Let us begin with the most general spacetimes conformal to static ones. In advanced coordinates, their metric takes the form
\begin{equation}
    ds^2= e^{\Lambda(v,r)} \left(- e^{2 h_+(r)} f(r) dv^2+ 2 e^{h_+} dv dr+ r^2 d\Omega^2\right). \label{come1}
\end{equation}
This spacetime possesses a conformal Killing vector $K^\mu= \left(1, 0,0,0\right)$. Applying the conformal Killing equation~\eqref{ckv6} gives
$2 {\cal B}= \partial\Lambda/\partial v$. Following Ref.~\cite{dahal:2023}, we now make a realistic assumption that the spacetime is slowly evolving. More precisely, we retain all terms in the equations containing $\Lambda$ and its first time derivative. Higher derivatives and terms quadratic in the first derivative are considered negligible
\begin{equation}
    e^{\Lambda} = 1+ \frac{d\Lambda}{dv} v+ {\cal O} \left(\left(\frac{d\Lambda}{dv}\right)^2, \frac{d^2\Lambda}{dv^2}\right). \label{ape3}
\end{equation}
Here we have normalized $e^{\Lambda}$ (by keeping the appropriate function dependent on the radial coordinate inside the bracket of Eq.~\eqref{come1}) such that the first term in its expansion is unity. Under this approximation ${\cal B}= \mathrm{Const.}$, and we can calculate the Misner-Sharp mass of such metric to show its linear dependence on $v$. For the specific example of the linear Vaidya spacetime, the value of ${\cal B}$ associated with the conformal Killing vector of Eq.~\eqref{ckv7} is $-1$, indeed a constant.

\section{Differential variations} \label{app6}

\subsection{Simplifications of the second and third terms of Eq.~\eqref{vricci53}}

For the general metric of Eq.~\eqref{come1}, we can express the infinitesimal diffeomorphism as
\begin{equation}
    h_{\mu\nu}= e^{\Lambda} h_{0\mu\nu} \approx \left(1+ 2 {\cal B} v\right) h_{0\mu\nu}, \label{dvf1}
\end{equation}
where $h_{0\mu\nu}$ represents the diffeomorphism associated with the static spacetime. Here, we are using the approximation of Eq.~\eqref{ape3} for simplifications. We thus have
\begin{equation}
    h_\mu^{~~[\mu;\beta]}= {\cal B} \left(g^{\nu\beta} \frac{\partial v}{\partial x^\nu} h_{0\mu}^{~~~\mu}- g^{\nu\mu} \frac{\partial v}{\partial x^\nu} h_{0\mu}^{~~~\beta}\right)+ \left(1+ 2 {\cal B} v\right) h_{0\mu}^{~~~[\mu;\beta]}.
\end{equation}
As the Lie derivative of a pair of diffeomorphically related static metrics along $K^\mu$ vanishes, we have
\begin{equation}
    {\mathcal L_K} \left(h_\mu^{~~[\mu;\beta]}\right) \approx 2 {\cal B} h_\mu^{~~[\mu;\beta]}.
\end{equation}
Similarly, using Eq.~\eqref{ckv6}, we get
\begin{equation}
    K^\nu_{~~;\nu} h_\mu^{~~[\mu;\beta]}= 4 {\cal B} h_\mu^{~~[\mu;\beta]}.
\end{equation}
We thus obtain
\begin{equation}
    -\frac{1}{4\pi} \int_\Sigma \left({\mathcal L_K} \left(h_\mu^{~~[\mu;\beta]}\right)+ K^\nu_{~~;\nu} h_\mu^{~~[\mu;\beta]}\right) {\hat T}_\beta \sqrt{|\gamma^{(\Sigma)}|} d^3 x= -\frac{1}{4\pi} \int_\Sigma 6 {\cal B} h_\mu^{~~[\mu;\beta]} {\hat T}_\beta \sqrt{|\gamma^{(\Sigma)}|} d^3 x. \label{vf5}
\end{equation}
As ${\hat T}^\gamma$ denotes a timelike unit vector normal to $\Sigma$ and ${\cal B} h_\mu^{~~\mu;\beta}$ is independent of time in the leading order, $h_\mu^{~~\mu;\beta} {\hat T}_\beta$ vanishes in that order. Consequently, one of the integral on the right hand side of this equation vanishes, thereby giving
\begin{equation}
    -\frac{1}{4\pi} \int_\Sigma \left({\mathcal L_K} \left(h_\mu^{~~[\mu;\beta]}\right)+ K^\nu_{~~;\nu} h_\mu^{~~[\mu;\beta]}\right) {\hat T}_\beta \sqrt{|\gamma^{(\Sigma)}|} d^3 x= \frac{1}{4\pi} \int_\Sigma 3 {\cal B} h_\mu^{~~\beta;\mu} {\hat T}_\beta \sqrt{|\gamma^{(\Sigma)}|} d^3 x. \label{lf6}
\end{equation}
Now, let us calculate
\begin{align}
    h_\mu^{~~\beta;\mu} {\hat T}_\beta=& \left(h_\mu^{~~\beta} {\hat T}_\beta\right)^{;\mu}- h_\mu^{~~\beta} {\hat T}_\beta^{~~;\mu}= \left(h_{\mu\beta} {\hat T}^\beta\right)^{;\mu}- h^{\mu\beta} {\hat T}_{\beta;\mu} \nonumber\\
    =& \left(\delta\left(g_{\mu\beta} {\hat T}^\beta\right)- g_{\mu\beta} \delta{\hat T}^\beta \right)^{;\mu}- \delta\left(-{\hat T}^\mu {\hat T}^\beta+ {N}^\mu {N}^\beta+ \sigma^{\mu\beta}\right) {\hat T}_{\beta;\mu} \nonumber\\
    =& 2 \left(\delta{\hat T}_\mu\right)^{;\mu}- 2 \delta N^\mu N^\beta {\hat T}_{\beta;\mu}- \delta\sigma^{\mu\beta} {\hat T}_{\beta;\mu}= k_1 \left({\hat T}_\mu\right)^{;\mu}+ k_2 N^\mu N^\beta {\hat T}_{\beta;\mu}- \delta\sigma^{\mu\beta} {\hat T}_{\beta;\mu}, \label{hf7}
\end{align}
where $\sigma^{\mu\beta}$ represents the induced metric on the two-surface $\partial\Sigma$. For simplifications, we have used the fact that $\delta{\hat T}_\mu$ is parallel to ${\hat T}_\mu$ and $\delta N^\mu$ is parallel to $N^\mu$
\begin{equation}
    \delta{\hat T}_\mu= \frac{k_1}{2} {\hat T}_\mu, \qquad \delta N_\mu= \frac{k_2}{2} N_\mu, \label{df8}
\end{equation}
where $k_1$ and $k_2$ are proportionality factors. Moreover, we have also used the relation
\begin{align}
    \delta{\hat T}^\mu=& \delta\left( g^{\mu\beta}{\hat T}_\beta\right)= -g^{\mu\alpha} h_{\alpha\beta} {\hat T}^\beta+ \frac{k_1}{2} {\hat T}^\mu \nonumber\\
    =& -g^{\mu\alpha} \delta\left(-{\hat T}_\alpha {\hat T}_\beta+ {N}_\alpha {N}_\beta+ \sigma_{\alpha\beta}\right) {\hat T}^\beta+ \frac{k_1}{2} {\hat T}^\mu \nonumber\\
    =& 2 g^{\mu\alpha} \delta{\hat T}_\alpha {\hat T}_\beta {\hat T}^\beta+ \frac{k_1}{2} {\hat T}^\mu= -\frac{k_1}{2} {\hat T}^\mu,
\end{align}
and the analogous expression for $\delta N^\mu$. In this calculation, we used the fact that $\delta \sigma_{\alpha\beta} {\hat T}^\beta= 0$ as $\delta \sigma_{\alpha\beta}\in \mathrm{span} \{U\otimes U, U\otimes V, V\otimes U, V\otimes V\}$ for some linearly independent vectors $U$ and $V$ generating two-surface $\partial\Sigma$. By this same reasoning, we also get $\delta\sigma^{\mu\beta} {\hat T}_{\beta;\mu}=0$, which simplifies Eq.~\eqref{hf7} to the form
\begin{equation}
    h_\mu^{~~\beta;\mu} {\hat T}_\beta= k_1 \left({\hat T}_\mu\right)^{;\mu}+ k_2 N^\mu N^\beta {\hat T}_{\beta;\mu}. \label{hf10}
\end{equation}
Furthermore, as $\sigma_{\alpha\beta}\in \mathrm{span} \{U\otimes U, U\otimes V, V\otimes U, V\otimes V\}$, we also have $\sigma^{\mu\beta} {\hat T}_{\mu;\beta}=0$, from which we obtain
\begin{equation}
    \left(g^{\alpha\beta}+ {\hat T}^\alpha {\hat T}^\beta- {N}^\alpha {N}^\beta\right) {\hat T}_{\alpha;\beta}=0 \implies {\hat T}^\beta_{~~;\beta}- N^\alpha N^\beta {\hat T}_{\alpha;\beta}= 0.
\end{equation}
This expression further simplifies Eq.~\eqref{hf10} to the form
\begin{equation}
    h_\mu^{~~\beta;\mu} {\hat T}_\beta= (k_1 + k_2) \left({\hat T}^\mu\right)_{;\mu}.
\end{equation}
Substituting this result into Eq.~\eqref{lf6}, we obtain
\begin{equation}
    -\frac{1}{4\pi} \int_\Sigma \left({\mathcal L_K} \left(h_\mu^{~~[\mu;\beta]}\right)+ K^\nu_{~~;\nu} h_\mu^{~~[\mu;\beta]}\right) {\hat T}_\beta \sqrt{|\gamma^{(\Sigma)}|} d^3 x= \frac{1}{4\pi} \int_\Sigma 3 {\cal B} (k_1 + k_2) \left({\hat T}^\mu\right)_{;\mu} \sqrt{|\gamma^{(\Sigma)}|} d^3 x.
\end{equation}
Now, we can use Gauss' theorem to convert the volume integral over $\Sigma$ into the surface integral over ${\partial\Sigma_\infty}$ and ${\partial\Sigma_{+}}$ bounding the volume $\Sigma$
\begin{multline}
    -\frac{1}{4\pi} \int_\Sigma \left({\mathcal L_K} \left(h_\mu^{~~[\mu;\beta]}\right)+ K^\nu_{~~;\nu} h_\mu^{~~[\mu;\beta]}\right) {\hat T}_\beta \sqrt{|\gamma^{(\Sigma)}|} d^3 x \\
    =  \frac{1}{4\pi} \int_{\partial\Sigma_\infty} 3 {\cal B} \left(k_1+ k_2\right) {\hat T}^\mu N_\mu \sqrt{|\gamma^{(\partial\Sigma_\infty)}|} d^2 y- \frac{1}{4\pi} \int_{\partial\Sigma_{+}} 3 {\cal B} \left(k_1+ k_2\right) {\hat T}^\mu N_\mu \sqrt{|\gamma^{(\partial\Sigma_{+})}|} d^2 y.
\end{multline}
At $\partial\Sigma_\infty$, ${\hat T}^\mu N_\mu = 0$ and at $\partial\Sigma_{+}$, ${\hat T}^\mu N_\mu = -1$, thereby giving
\begin{equation}
    -\frac{1}{4\pi} \int_\Sigma \left({\mathcal L_K} \left(h_\mu^{~~[\mu;\beta]}\right)+ K^\nu_{~~;\nu} h_\mu^{~~[\mu;\beta]}\right) {\hat T}_\beta \sqrt{|\gamma^{(\Sigma)}|} d^3 x= \frac{1}{4\pi} \int_{\partial\Sigma_{+}} 3 {\cal B} \left(k_1+ k_2\right) \sqrt{|\gamma^{(\partial\Sigma_{+})}|} d^2 y. \label{lf15}
\end{equation}

\subsection{Differential variations of the surface gravity} \label{app62}

Unlike previous calculations, we constrain ourselves at the horizon in this subsection, where both the conformal Killing vector and ${\hat T}^\mu$ becomes null. Starting from Eq.~\eqref{sg45}, which defines $\kappa$, we can express its variation as
\begin{equation}
    \delta\kappa= \delta\left(\left(K_{\mu;\beta}- 2 {\cal B} g_{\mu\beta}\right) N^\mu {\hat T}^\beta\right).
\end{equation}
However, from Eq.~\eqref{sg40}, we have $K_{\mu;\beta}- 2 {\cal B} g_{\mu\beta}= -K_{\beta;\mu}$. Applying this relation
\begin{equation}
    \delta\kappa= -\delta\left(K_{\beta;\mu} N^\mu {\hat T}^\beta\right)= -\frac{1}{2} \delta\left( \left({\hat T}_\beta {\hat T}^\beta\right)_{;\mu} N^\mu \right),
\end{equation}
where last equality utilizes the fact that at the horizon, $K^\beta={\hat T}^\beta$. Using the product rule for differentiation yields
\begin{equation}
    \delta\kappa= -\frac{1}{2} \left(\delta {\hat T}_{\beta} {\hat T}^\beta+ {\hat T}_\beta \delta {\hat T}^\beta \right)_{;\mu} N^\mu -\frac{1}{2} \left({\hat T}_\beta {\hat T}^\beta\right)_{;\mu} \delta N^\mu.
\end{equation}
We now use the fact that $\delta {\hat T}^\beta=0$ to simplify the expression for $\delta\kappa$
\begin{equation}\begin{aligned} \label{sg60}
    \delta\kappa=& -\frac{1}{2} \left(\delta {\hat T}_{\beta;\mu} {\hat T}^\beta N^\mu+ \delta {\hat T}_\beta {\hat T}^\beta_{~~;\mu} N^\mu \right) - {\hat T}_\beta {\hat T}^\beta_{~~;\mu} \delta N^\mu \\
    =& -\frac{1}{2} \delta {\hat T}_{\beta;\mu} \left({\hat T}^\beta N^\mu+ {\hat T}^\mu N^\beta \right)+ \frac{1}{2} \left(\delta {\hat T}_{\beta;\mu} {\hat T}^\mu N^\beta- \delta {\hat T}_\beta {\hat T}^\beta_{~~;\mu} N^\mu\right) - {\hat T}_\beta {\hat T}^\beta_{~~;\mu} \delta N^\mu.
\end{aligned}\end{equation}
Furthermore, $\delta {\hat T}_{\beta}$ is proportional to ${\hat T}_{\beta}$, yielding the relation
\begin{equation}
    \delta {\hat T}_{\beta;\mu}  {\hat T}^{\mu}- {\hat T}_{\beta;}^{~~~\mu} \delta {\hat T}_{\mu}= 0.
\end{equation}
By substituting this relationship into the conformal Killing equation, we obtain
\begin{equation}
    \delta {\hat T}_{\beta;\mu}  {\hat T}^{\mu}= - {\hat T}^\mu_{~~;\beta}  \delta{\hat T}_{\mu}+ 2 {\cal B} g^\mu_{~~\beta} \delta{\hat T}_{\mu}.
\end{equation}
This relation, when substituted back into Eq.~\eqref{sg60} yields
\begin{equation}
    \delta\kappa= -\frac{1}{2} \delta {\hat T}_{\beta;\mu} \left({\hat T}^\beta N^\mu+ {\hat T}^\mu N^\beta \right)+ \frac{1}{2} \left(- 2 {\hat T}^\mu_{~~;\beta} \delta{\hat T}_\mu N^\beta+ 2 {\cal B} g^\mu_{~~\beta} \delta{\hat T}_{\mu} N^\beta \right) - {\hat T}_\beta {\hat T}^\beta_{~~;\mu} \delta N^\mu.
\end{equation}
Both terms ${\hat T}_\beta {\hat T}^\beta_{~~;\mu} \delta N^\mu$ and ${\hat T}^\mu_{~~;\beta} \delta{\hat T}_\mu N^\beta$ vanish (the second one because $\delta {\hat T}_{\beta}$ is proportional to ${\hat T}_{\beta}$), thereby leading the simplified expression
\begin{equation}
    \delta\kappa= -\frac{1}{2} \delta {\hat T}_{\beta;\mu} \left({\hat T}^\beta N^\mu+ {\hat T}^\mu N^\beta \right)+ {\cal B} g^\mu_{~~\beta} h_{\mu\gamma} {\hat T}^\gamma N^\beta .
\end{equation}
Moreover, $\delta {\hat T}_{\beta}$ is proportional to ${\hat T}_{\beta}$ on the horizon also implies that $\left(\delta {\hat T}_\alpha\right)_{;\beta} m^\alpha \tilde m^\beta= 0$ in spherically symmetric spacetimes, allowing us to write the variation of surface gravity equivalently as
\begin{equation}
    \delta\kappa= \frac{1}{2} \left( \delta {\hat T}_{\beta} \right)^{;\beta}+ {\cal B} h_{\beta\gamma} {\hat T}^\gamma N^\beta = \frac{1}{2} \left( h_{\beta\gamma} \right)^{;\beta} {\hat T}^\gamma+ {\cal B} h_{\beta\gamma} {\hat T}^\gamma N^\beta. \label{sgf24}
\end{equation}
Here, $m^\beta$ is a null vector on two-surface $\partial\Sigma$ and $\tilde m^\beta$ is its complex conjugate. This equation allows us to express the following integral into a desired form
\begin{multline}
    \frac{1}{4\pi} \int_{\partial\Sigma_{+}} \left(K^\beta h_\mu^{~~[\mu;\nu]}- K^\nu h_\mu^{~~[\mu;\beta]}\right) N_\nu {\hat T}_\beta \sqrt{|\gamma^{(\partial\Sigma_{+})}|} d^2 y= \frac{1}{4\pi} \int_{\partial\Sigma_{+}} h_\mu^{~~[\mu;\beta]} {\hat T}_\beta \sqrt{|\gamma^{(\partial\Sigma_{+})}|} d^2 y \\
    = \frac{1}{4\pi} \int_{\partial\Sigma_{+}} \left(4 {\cal B} k_0- 2 {\cal B} (k_3+ k_6) -\delta\kappa+ {\cal B} h_{\beta\gamma} {\hat T}^\gamma N^\beta\right) \sqrt{|\gamma^{(\partial\Sigma_{+})}|} d^2 y,
\end{multline}
where we have used relations $K^\beta {\hat T}_\beta=0$ and $K^\nu N_\nu= -1$ at the event horizon to write the first equality. To write second equality, we have used Eq.~\eqref{sgf24} and the relation
\begin{equation}
    h_\mu^{~~\mu;\beta} {\hat T}_\beta= 4 {\cal B} k_0- 2 {\cal B} (k_3+ k_6). \label{dvf26}
\end{equation}
To establish this equality, we substitute Eq.~\eqref{dvf1} into its left hand side
\begin{equation}
    h_\mu^{~~\mu;\beta} {\hat T}_\beta= \frac{\partial}{\partial x^\alpha} \left(\left(1+ 2 {\cal B} v\right) h_{0\mu}^{~~~\mu} \right) {\hat T}^\alpha \approx 2 {\cal B} h_{\mu}^{~~\mu}. \label{dvf27}
\end{equation}
Now, it remains to calculate the trace of the infinitesimal diffeomorphism $h_{\mu}^{~~\mu}$. For this, we begin from the condition $g^{\mu\nu} {\hat T}_\mu N_\nu= -1$ that holds on $\partial\Sigma_{+}$, which implies
\begin{equation}
     \delta g^{\mu\nu} {\hat T}_\mu N_\nu+ g^{\mu\nu} \delta{\hat T}_\mu N_\nu+ g^{\mu\nu} {\hat T}_\mu \delta N_\nu= 0. \label{mv60}
\end{equation}
Substituting $\delta{\hat T}_\mu$ and $\delta N_\nu$ from Eq.~\eqref{df8}, we obtain
\begin{equation}
    k_1+ k_2= -2 h^{\mu\nu} {\hat T}_\mu N_\nu \eqdef 2 k_0, \label{dv61}
\end{equation}
Following the exact same procedure, we obtain
\begin{equation}
    h^{\mu\nu} {\hat T}_\mu {\hat T}_\nu = 0, \quad h^{\mu\nu} N_\mu N_\nu= 0.
\end{equation}
Next, we consider the remaining two vectors $U$ and $V$, which generate the two-surface $(\partial\Sigma_{+})$ and their most general differential perturbation can be expressed as
\begin{equation}
    \delta U_\mu= \frac{1}{2} k_3 U_\mu+ \frac{1}{2} k_5 V_\mu, \quad \delta V_\mu= \frac{1}{2} k_4 U_\mu+ \frac{1}{2} k_6 V_\mu. \label{f30}
\end{equation}
Applying the same procedure as before and then employing these relations, we obtain
\begin{equation}
    h^{\mu\nu} U_\mu U_\nu = -k_5, \quad h^{\mu\nu} U_\mu V_\nu = -\frac{k_3+ k_6}{2}, \quad h^{\mu\nu} V_\mu V_\nu = -k_4.
\end{equation}
These differential variations in $h^{\mu\nu}$ can be converted from the tetrad frame to the tensor frame, yielding
\begin{equation}
    h_\mu^{~~\mu}= 2 k_0- (k_3+ k_6).
\end{equation}
Substituting this relation back into Eq.~\eqref{dvf27} leads to Eq.~\eqref{dvf26}.

\end{widetext}

\newpage{\pagestyle{empty}\cleardoublepage}

\end{document}